\documentclass[prb,aps,showpacs,superscriptaddress,twocolumn]{revtex4-1}
\usepackage{latexsym}
\usepackage{amsfonts}
\usepackage{amssymb}
\usepackage{amsmath}
\usepackage{graphicx}
\begin{document}
\title{Spin-polarized tunneling spectroscopic studies of the intrinsic\\
heterogeneity and pseudogap phenomena in colossal\\
magnetoresistive manganite $\rm La_{0.7}Ca_{0.3}MnO_3$}
\date{\today}
\author{C.~R. Hughes,$^1$ J. Shi,$^1$ A. D. Beyer,$^1$ and N.-C. Yeh$^{2,}$}

\affiliation{Department of Physics, California Institute of Technology, Pasadena, CA 91125 \\ $^2$Kavli Nanoscience Institute, California Institute of Technology, Pasadena, CA 91125}

\begin{abstract}
Spatially resolved tunneling spectroscopic studies of colossal magnetoresistive (CMR) manganite $\rm La_{0.7}Ca_{0.3}MnO_3$ (LCMO) epitaxial films on $\rm (LaAlO_3)_{0.3}(Sr_2AlTaO_6)_{0.7}$ substrate are investigated as functions of temperature, magnetic field and spin polarization by means of scanning tunneling spectroscopy. Systematic surveys of the tunneling spectra taken with Pt/Ir tips reveal spatial variations on the length scale of a few hundred nanometers in the ferromagnetic state, which may be attributed to the intrinsic heterogeneity of the manganites due to their tendency towards phase separation. The electronic heterogeneity is found to decrease either with increasing field at low temperatures or at temperatures above all magnetic ordering temperatures. On the other hand, spectra taken with Cr-coated tips are consistent with convoluted electronic properties of both LCMO and Cr. In particular, for temperatures below the magnetic ordering temperatures of both Cr and LCMO, the magnetic-field dependent tunneling spectra may be quantitatively explained by the scenario of spin-polarized tunneling in a spin-valve configuration. Moreover, a low-energy insulating energy gap $\sim 0.6$ eV commonly found in the tunneling conductance spectra of bulk metallic LCMO at $T \to 0$ may be attributed to a surface ferromagnetic insulating (FI) phase, as evidenced by its spin filtering effect at low temperatures and vanishing gap value above the Curie temperature. Additionally, temperature independent pseudogap (PG) phenomena existing primarily along the boundaries of magnetic domains are observed in the zero-field tunneling spectra. The PG becomes strongly suppressed by applied magnetic fields at low temperatures when the tunneling spectra of LCMO become highly homogeneous. These findings suggest that the occurrence PG is associated with the electronic heterogeneity of the manganites. The observation of lateral and vertical electronic heterogeneity in the CMR manganites places important size constraints on the development of high-density nano-scale spintronic devices based on these materials.  
\end{abstract}
\pacs{75.47.Lx, 73.40.Gk, 72.25.-b, 71.27.+a} \maketitle

\section{Introduction}

``Spintronics'' is a new paradigm of electronics based on the spin-dependent charge transport in magnetic heterostructures.~\cite{Wolf01,GuptaJA01} It has emerged as one of the most active research fields in recent years because of the potential advantages of non-volatility, faster processing speed, small power dissipation for high-density device integration as opposed to conventional semiconducting systems,~\cite{Wolf01} and better coherence promising for quantum information technology.~\cite{GuptaJA01} 

Among the physics issues associated with spintronics, knowledge of spin-polarized quantum tunneling and transport across interfaces is particularly important for developing high quality and reproducible spintronic devices. From the perspective of materials, ferromagnets with higher degrees of spin polarization and higher Curie temperatures are desirable candidates for use in spintronic devices. In this context, the manganese oxides $\rm Ln_{1-x}A_xMnO_3$ (Ln: trivalent rare earth ions, A: divalent alkaline earth ions), also widely known as manganites that exhibit colossal magnetoresistance (CMR) effects,~\cite{Kusters89,vonHelmolt93,Jin94,Xiong95,Ramirez97,Coey99} appear to be promising spintronic materials because of their half-metallicity in the ferromagnetic state~\cite{Dagotto01,Pickett96,Satpathy96} so that the degree of spin polarization is nearly 100\%. Nonetheless, experimental data and theoretical calculations have suggested that the ground states of the manganites tend to be intrinsically inhomogeneous as the result of their strong tendencies toward phase separation, and the phase separation may involve domains of ferromagnetic metals, ferromagnetic insulators, and antiferromagnetic charge and orbital ordered insulators.~\cite{Dagotto01,Fath99,Renner00,Becker02,Algarabel03,Aruta09} In fact, it has been demonstrated numerically that the double exchange interaction and half-metallicity alone cannot account for the large magnitude of negative magnetoresistance in ferromagnetic manganites,~\cite{Dagotto01} and that the tendencies toward phase separation in the ground state even for the nominally metalltic ferromagnetic phases play an essential role in the occurrence of the CMR effects.~\cite{Dagotto01} From the viewpoint of technological applications, the intrinsic electronic heterogeneity of the manganites becomes a relevant concern for consistently fabricating miniaturized spintronic devices with high areal densities. In this context, understanding of the spatial varying physical characteristics of the manganites at the microscopic scale will be important for spintronic applications based on these materials. 

In addition to the tendency toward phase separation in bulk manganites, various experimental findings have suggested that the surfaces of manganites appear to differ from the bulk characteristics.~\cite{Aruta09} For instance, scanning tunneling spectroscopic (STS) studies of nominally metallic manganite epitaxial thin films~\cite{Wei97,Wei98,Seiro07,Moshnyaga06} and single crystals~\cite{Renner00} have always revealed either a small energy gap or a pseudogap near the Fermi surface. Although the occurrence of an energy gap in a nominally metallic manganite could not be explained by either bandstructure calculations~\cite{Pickett96,Satpathy96} or bulk electrical transport measurements,~\cite{Yeh97,Yeh97a} the tunneling spectra of the metallic manganite over a wide energy range except near the Fermi level were actually consistent with theoretical calculations of the bulk electronic density of states.~\cite{Wei97,Wei98} Hence, it is reasonable to conjecture that the surface of manganites may be a thin  layer with chemical compositions differing from those of the bulk. Thus, lower-bias ballistic electrons injected from the scanning tunneling microscopy (STM) tip could be more sensitive to the surface state, whereas ballistic electrons injected with higher-bias could penetrate deeper into the bulk. In other words, the spectroscopic characteristics at higher bias voltages may be more representative of the bulk density of states, whereas those at lower bias voltages may be better related to the surface state. This conjecture is consistent with the x-ray photoemission spectroscopic (XPS) studies of $\rm La_{0.67}Ca_{0.33}MnO_3$ that inferred a surface predominantly terminated by an insulating layer of $\rm MnO_2$.~\cite{Broussard97} 

Another interesting feature associated with the manganites is the occurrence of pseudogap (PG) phenomena.~\cite{Dagotto01} Theoretical studies using Monte Carlo simulations~\cite{Moreo99,Moreo00} suggest that the density of states (DOS) in the manganites should exhibit PG characteristics, with significant spectral depletion at the chemical potential and broad DOS peaks both above and below the chemical potential. These theoretical findings have been corroborated by photoemission experiments on bilayer manganites above their magnetic ordering temperatures.~\cite{Dessau98} Theoretically, the occurrence of PG may be regarded as a precursor of phase separations in forms of magnetic clusters,~\cite{Moreo99,Moreo00} which is analogous to the widely studied PG phenomena in cuprate superconductors where the appearance of PG is attributed to the onset of preformed pairs and competing orders.~\cite{Yeh09,Yeh10} However, whether the PG phenomena are common among different types of manganites and whether the physical origin of PG in the manganites is indeed associated with the onset of mixed phases have not been extensively verified by experiments.  

To address the aforementioned issues of phase separations, insulating surface layers and the PG phenomena in the manganites, we report in this work spatially resolved tunneling spectroscopic studies of $\rm La_{0.7}Ca_{0.3}MnO_3$ (LCMO) epitaxial films by means of both regular and spin-polarized scanning tunneling microscopy~\cite{Bode03,Czerner09,Alvarado95,Alvarado92,Meier08} (STM and SP-STM). The specific calcium doping level of $x = 0.3$ was chosen because it corresponded to a nominal metallic phase with nearly the highest Curie temperature ($T_C \sim 270$ K) and the most spatial homogeneity among the Ca-doped manganites. The evolution of the spatially resolved tunneling spectra was studied systematically with temperature, magnetic field and the degree of spin polarization, which provided information about the spatial scales of stoichiometric inhomogeneity and the average size of ferromagnetic domains. Additionally, comparison with bandstructure calculations suggested that the spectral characteristics taken with regular STM were consistent with those of the DOS of the manganite, whereas data taken with SP-STM may be understood in terms of the product of a spin-dependent tunneling matrix and the joint density of states between the SP-STM tip material and the manganite. on the other hand, the evolution of the surface energy gap with temperature, magnetic field and the degree of spin-polarization was found to be consistent with the spin filtering effect of a surface ferromagnetic insulating (FI) phase.~\cite{Algarabel03} Finally, PG phenomena were found to persist at temperatures well above all magnetic ordering temperatures in the absence of external magnetic fields. The PG features were suppressed by moderate magnetic fields at low temperatures when the tunneling conductance of LCMO became spatially homogeneous. 

\begin{figure}
  \centering
  \includegraphics[width=3.4in]{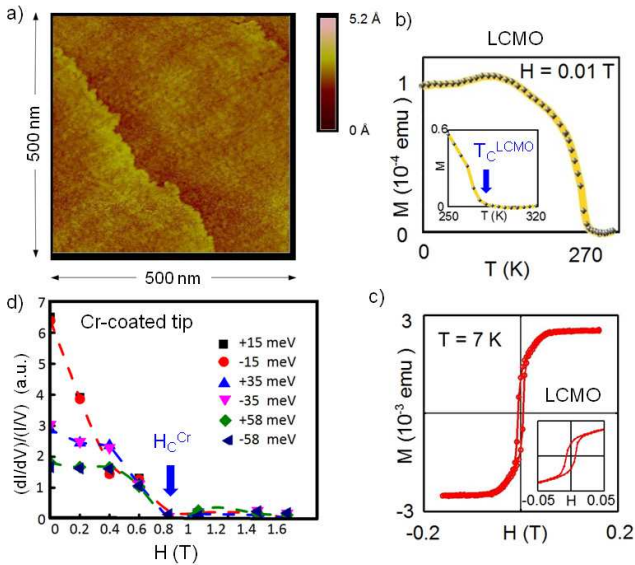}
\caption{(Color online) (a) Typical STM topography of a $\rm La_{0.7}Ca_{0.3}MnO_3$ epitaxial film on $\rm (LaAlO_3)_{0.3}(Sr_2AlTaO_6)_{0.7}$ substrate over a $(500 \times 500) \rm nm^2$ sample area, showing atomically flat surfaces and steps of one lattice constant height. (b) Magnetization ($M$) vs. temperature ($T$) data of the same LCMO epitaxial film taken under an applied magnetic field $H = 0.01$ T, showing $T_C ^{\rm LCMO} = 270 \pm 10$ K with detailed temperature dependence near $T_C ^{\rm LCMO}$ given in the inset. (c) Main panel: $M$-vs.-$H$ hysterisis curve of the same LCMO film taken at a constant temperature $T$ = 7 K and for $H$ from $-0.2$ T to 0.2 T. Inset: The same $M(H)$ curve as in the main panel and for $-0.05 {\rm T} \le H \le 0.05 {\rm T}$, showing a LCMO coercive field $H_C ^{\rm LCMO} \sim 0.02$ T. (d) Normalized tunneling conductance vs. $H$ spectra obtained from tunneling electrical currents from a Cr-coated tip to a permanent magnet, a NiNd single crystal, at constant bias voltages $V = \pm 15$, $\pm 35$ and $\pm 58$ mV and for $T = 6$ K. The magnetic polarization of NiNd was opposite to the applied magnetic field. Therefore, when the applied field exceeded the coercive field of the Cr-coated tip, the spin-polarized tunneling from the Cr-coated tip to NiNd became forbidden near the Fermi level, yielding nearly zero tunneling conductance. The coercive field thus determined is $|H_{C} ^{\rm Cr}| = 0.8 \pm 0.1$ T.}
\label{fig1}
\end{figure}

The findings presented in this work provide quantitative experimental accounts for the presence of phase separations, surface states and PG phenomena in the manganites. In comparison with other STM studies of the manganites, while previous investigations have revealed spatially inhomogeneous spectra in manganites with different doping levels,~\cite{Fath99,Renner00,Becker02} this work provides the first high-field SP-STM studies on the manganites and demonstrates the general application of high-field SP-STM techniques to the investigation of spatially inhomogeneous magnetic materials. 

\section{Experimental}

The $\rm La_{0.7}Ca_{0.3}MnO_3$ (LCMO) films used in this study were epitaxially grown on $\rm (LaAlO_3)_{0.3}(Sr_2AlTaO_6)_{0.7}$ substrates by means of pulsed laser deposition (PLD) techniques. The substrates were chosen because of their small lattice mismatch (about 0.3\%) with the bulk LCMO, which ensures minimized strain in the resulting films to prevent excess strain-induced effects on the electronic properties.~\cite{Yeh97,Yeh97a,Mitra05} The films were deposited in a 100 mTorr oxygen background pressure to a thickness of $(110 \pm 10)$ nm with the substrate temperature kept at 650$^{\circ}$C, and were subsequently annealed at the same temperature for 2 hours in 100 Torr O$_2$ and then cooled slowly to room temperature. The epitaxy of these films was confirmed by x-ray diffraction, and the film quality was further verified by atomic force microscopy and STM measurements, showing terraced growth with step heights corresponding to one c-axis lattice constant of the bulk LCMO, as exemplified in Fig.~\ref{fig1}(a). Further characterizations were conducted using a Superconducting-Quantum-Interference-Device (SQUID) magnetometer by Quantum Design for magnetization measurments, showing a Curie temperature of $270 \pm 10$ K as exemplified in Fig.~\ref{fig1}(b) for one of the LCMO epitaxial films. Additionally, the coercive fields of the LCMO films ($H_{C}^{\rm LCMO}$) for magnetic fields perpendicular to the plane of films were determined by the SQUID magnetometer, and were found to range from 0.02 T to 0.04 T as exemplified in Fig.~\ref{fig1}(c). In general, LCMO thin films fabricated under aforementioned conditions have been shown to yield high-quality epitaxy via high-resolution x-ray diffraction studies, magnetization and transport measurements.~\cite{Yeh97,Yeh97a,Yeh97b} 

Prior to STM measurements, samples were first etched in a 0.5\% bromine in pure ethanol solution and then rinsed in pure ethanol to remove surface contaminants such as the carbonates. Each etched sample was immediately loaded onto our cryogenic STM system while kept under excess pressure of helium gas during the loading process. It is worth noting that the presence of surface contaminants is common among the perovskite oxides that contain alkaline earth metals (A = Ca, Sr and Ba) because the alkaline earth elements are highly reative to H$_2$O and CO$_2$.~\cite{Vasquez94} Consequently, the removal of surface non-stoichiometric alkaline earth compounds such as AO, ACO$_3$ and A(OH)$_2$ upon chemical etching and ethonal rinsing would tend to yield slight Ca-deficiency on the sample surface relative to its bulk stoichiometry unless long-time etching was carried out.~\cite{Vasquez94} Further, sample surfaces without proper removal of the contaminants often exhibit excess concentrations of alkaline earth metals in the XPS studies.~\cite{Vasquez94,ChoiJ99} 
	
The tunneling studies were conducted with our homemade cryogenic STM with a base temperature $T = 6$ K and a superconducting magnet capable of magnetic fields up to $H = 7$ T. At $T = 6$ K the STM system was under ultra-high vacuum with a base pressure $\sim 10^{-10}$ mbar. Regular tunneling spectroscopic studies were conducted with atomically sharp Pt/Ir tips, while spin polarized studies were made by means of Pt/Ir tips evaporatively coated with 15 $\sim$ 30 monolayers of Cr metal prepared in a separate evaporation system, following the procedures reported previously.~\cite{Bode03,Czerner09} Here we note that thin-film Cr is ferromagnetic,~\cite{Bode03,Czerner09} and the Curie temperature ($T_C ^{\rm Cr}$) of our Cr-coated tip was found to be much higher than room temperature. Therefore, it is justifiable to assume that electrical currents from the Cr-coated tip were always spin-polarized. This assumption was subsequently verified by field-dependent tunneling conductance measurements, and the degree of spin polarization ($\sim 15$\%) was also estimated, as elaborated in Section III. The coercive field ($H_C ^{\rm Cr}$) of the Cr-coated tips was empirically determined by studying the tunneling conductance vs. magnetic field $H$ for electrical currents tunneling from the Cr-coated tip to a permanent magnet NdNi. The magnetization of NdNi was opposite to the applied magnetic field, so that for fields exceeding the coercive field of the Cr-coated tip, tunneling from the spin-polarized tip to NdNi became forbidden near the Fermi level and the tunneling conductance approached zero, as shown in Fig.~\ref{fig1}(d) for small bias voltages at $V = \pm 15$, $\pm 35$, and $\pm 58$ mV. The $H_C ^{\rm Cr}$ value thus determined was $0.8 \pm 0.1$ T, comparable to the values determined by similar tunneling conductance measurements outlined in Refs.~\onlinecite{Bode03,Czerner09}. Thus, we have established the condition $H_C ^{\rm LCMO} \ll H_C ^{\rm Cr} < 1.0$ T for all Cr-coated tips. 

\begin{figure*}
  \centering
  \includegraphics[width=6.0in]{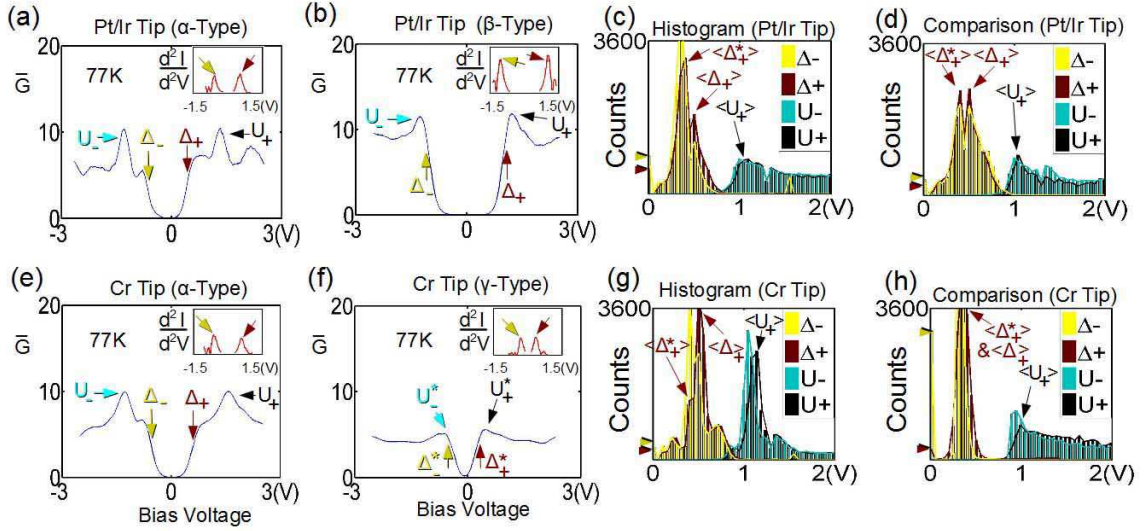}
\caption{(Color online) Characteristic features in the regular and spin-polarized tunneling spectra of LCMO taken at $T = 77$ K and $H = 0$: (a) A normalized $\alpha$-type tunneling conductance spectrum taken with a Pt/Ir tip, showing maximal conductance at $\omega = -U_-$ and $\omega = U_+$ that corresponds to the maximal DOS associated with the majority and minority bands of LCMO, respectively.~\cite{Pickett96} Additional features associated with an insulating energy gap at $\omega = -\Delta_-$ and $\omega = \Delta_+$ are observed. Here the values of $\Delta _{\pm}$ are determined by identifying the energies where the second derivatives of the tunneling current ($I$) relative to the bias voltage ($V$) reach maxima, as indicated in the inset. (b) A normalized $\beta$-type tunneling conductance spectrum taken with a Pt/Ir tip, showing one set of conductance peaks at $\omega = -U_-$ and $\omega = U_+$. The insulating gap values $\Delta _{\pm}$ are determined in the same way as in (a) and are shown in the inset. (c) The histograms of the characteristic energies $U_{\pm}$ and $\Delta_{\pm}$ derived from the tunneling spectra taken with a Pt/Ir tip over a $(500 \times 500) \ \rm nm^2$ sample area. Here $\langle U_+ \rangle$ refers to the maximal count of the positive characteristic energy that is associated with the DOS peak in LCMO, and $\Delta_+ ^{\ast}$ refers to the positive PG energy associated with the $\gamma$-type spectra. (d) Energy histograms of $U_{\pm}$ and $\Delta_{\pm}$ derived from the tunneling spectra taken with a Pt/Ir tip over a $(500 \times 500) \ \rm nm^2$ sample area different from the area studied in (c), showing statistically similar results. (e) A normalized $\alpha$-type tunneling conductance spectrum taken with a Cr-coated tip, showing similar features to the spectra taken with a Pt/Ir tip, although the values of $U_{\pm}$ and $\Delta _{\pm}$ differ slightly. (f) A normalized $\gamma$-type tunneling conductance spectrum taken with a Cr-coated tip, showing PG-like behavior with one set of relatively low conductance peaks at $\omega = -U_- ^{\ast}$ and $\omega = U_+ ^{\ast}$ and vanishing DOS at $\omega \sim 0$. Here we use the notation $\ast$ to represent features associated with the PG-like spectra. (g) Energy histograms of $U_{\pm}$ and $\Delta_{\pm}$ derived from the tunneling spectra taken with a Cr-coated tip over a $(500 \times 500) \ \rm nm^2$ sample area. (h) Energy histograms of $U_{\pm}$ and $\Delta_{\pm}$ derived from the tunneling spectra taken with a Cr-coated tip over a $(500 \times 500) \ \rm nm^2$ sample area different from that in (g). Here the values of the hard insulating gap and those of the PG appear to merge, whereas the distribution of $U_{\pm}$ appear to be comparable to that shown in (g).}
\label{fig2}
\end{figure*}

In the following studies, the STM system was operated at $T = 6$ K for $H = 0$, $-0.3$ T, and 3.0 T, and also at $T = 77$ K and 300 K for $H = 0$. Both Pt/Ir and Cr-coated tips were used in these studies. Therefore, the field-dependent studies were carried out at $T \ll T_C ^{\rm LCMO} \ll T_C ^{\rm Cr}$ and under three conditions: $H = 0$, $H_C^{\rm LCMO} < |H| < H_C^{\rm Cr}$ with $H < 0$, and $|H| > H_C^{\rm Cr}$ with $H > 0$. These conditions ensured that the effects of spin-polarized currents were investigated for three different magnetic configurations in LCMO, which are 1) randomly oriented magnetic domains, 2) aligned LCMO magnetic domains that were anti-parallel to the spin-polarized current, and 3) aligned LCMO magnetic domains that were parallel to the spin polarized current. 

The spectroscopic measurements consisted of tunneling current ($I$) versus bias voltage ($V$) spectra taken from $V = -3$ V to 3 V at each pixel on a $(128 \times 128)$ pixel grid located over a $\rm (500 \times 500) \ nm^2$ area for zero-field studies at 77 K and 300 K. On the other hand, the scanned sample area for field-dependent measurements at 6 K was reduced to $\rm (250 \times 90) \ nm^2$ because of the much reduced scanning range of the piezoelectric material at low temperatures and the experimental preference to complete a full spectroscopic scan without interruptions by the need of liquid helium transfer. For all tunneling spectra, the typical junction resistance was kept at $\sim 100 \rm M\Omega$. It was verified that the tunneling spectroscopy and topography were both independent of slight variations of the tip height relative to the sample surface under this range of junction resistance. For consistent analysis of all experimental data taken under different conditions on multiple samples, the tunneling spectra to be discussed below were all processed into normalized differential conductance, $(dI/dV)/(I/V) \equiv \bar{G}$, because this quantity minimizes extrinsic effects incurred by slight variations in the sample-tip separation, and so best represents the material characteristics of the tunnel junction. 

For comparison of the spectral characteristics as functions of temperature, magnetic field and spin polarization, it would have been ideal to conduct all studies over identical sample areas. However, in practice we were only able to investigate the field dependent spectra over the same sample area by keeping the measurements at a constant temperature $T = 6$ K. This limitation was because changes in the measurement temperature would result in drifts of the STM tip, and replacing the STM tip ($e.g.$ from Pt/Ir to Cr-coated tips) would lead to a different sample area upon reapproaching the tip to the sample. Given the intrinsic heterogeneity of LCMO, meaningful comparison among data taken either with different STM tips or at different temperatures could only be made if statistical consistency in the spectral characteristics could be established. This premise was indeed verified in our investigation, as elaborated further in Section III.

\section{Results and Analysis}

Based on the methods outlined in Section II, systematic studies of the tunneling spectral evolution with temperature, magnetic field and spin polarization were carried out to address the issues of phase separation, surface state and PG phenomena in the nominally metallic manganite epitaxial thin films of $\rm La_{0.7}Ca_{0.3}MnO_3$ (LCMO) on $\rm (LaAlO_3)_{0.3}(Sr_2AlTaO_6)_{0.7}$ substrates. 

\subsection{Spectral characteristics}

Our detailed surveys of the LCMO tunneling spectra over relatively large sample areas and multiple samples with both Pt/Ir and Cr-coated tips at $T = 77$ K ($\ll T_C ^{\rm LCMO} \ll T_C ^{\rm Cr}$) and $H = 0$ revealed three types of representative tunneling spectra, which are labeled as $\alpha$, $\beta$, and $\gamma$-types for convenience. Examples of the $\alpha$ and $\beta$-types of spectra taken with a Pt/Ir tip are shown respectively in Figs.~\ref{fig2}(a) and \ref{fig2}(b), and representative spectra of the $\alpha$ and $\gamma$-types taken with a Cr-coated tip are illustrated in Figs.~\ref{fig2}(e) and \ref{fig2}(f). 

For the dominant $\alpha$-type of spectra, there are four primary characteristic features, including two major conductance peaks at energies of $\omega = U_+$ and $- U_-$, and two smaller peaks flanking a low-energy insulating gap at the Fermi level. The $\alpha$-type of spectra is consistent with our previous single-point spectroscopic studies.~\cite{Wei97} In the case of the $\beta$-type spectra as exemplified in Fig.~\ref{fig2}(b), only one pair of peak features at $\omega = U_+$ and $\omega = - U_-$ may be identified, and this type of spectra typically reveals a wider tunneling gap. The $\gamma$-type spectra are essentially PG-like, as shown in Fig.~\ref{fig2}(f). Specifically, the characteristic energies $U_{\pm} ^{\ast}$ associated with the broad peak features of the $\gamma$-type spectra are much smaller than those found in the other two types of spectra, and the conductance values at $\omega = U_+ ^{\ast}$ and $\omega = -U_- ^{\ast}$ are also much smaller. Moreover, in contrast to the completely vanished density of states (DOS) over a finite range of energies in the $\alpha$ and $\beta$-type spectra, the DOS in the $\gamma$-type spectra only vanishes at the Fermi level and remains finite for all finite energies $\omega \not= 0$. 

\begin{figure}
  \centering
  \includegraphics[width=3.3in]{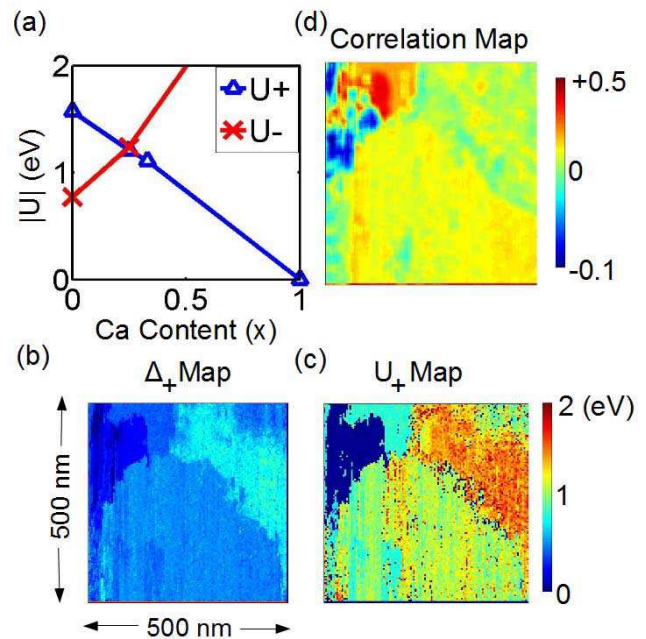}
\caption{(Color online) (a) Ca-doping ($x$) dependence of the DOS peak energies $U_{\pm}$ based on known bandstructure calculations,~\cite{Pickett96,Satpathy96} showing decreasing $U_+$ and increasing $U_-$ with increasing Ca-doping. Apparent correlation between the spatial distribution of the DOS peak energy $U_+$ and that of the surface insulating gap $\Delta_+$ is manifested by comparing the $\Delta _+$ map in (b) and the $U_+$ map (c), both over the same ($500 \times 500$) $\rm nm^2$ sample area. The correlation is further quantified by the map of cross correlation between $U_+$ and $\Delta _+$ in (d),~\cite{correlation} showing positive correlation throughout most of the sample area. A smaller fraction of anti-correlated regions occurs along the domain boundaries. These regions are associated with the occurrence of PG and exhibit the $\gamma$-type spectra.}
\label{fig3}
\end{figure}

For consistent comparison of the low-energy features among three different types of spectra, we associate $\Delta _+$ and $-\Delta _-$ (or $\Delta _+ ^{\ast}$ and $-\Delta _- ^{\ast}$ in the case of the $\gamma$-type) with the energies where the derivatives of the low-energy tunneling conductance reached the maximum, as shown in the insets of Figs.~\ref{fig2}(a), \ref{fig2}(b), \ref{fig2}(e) and \ref{fig2}(f). The histograms for all four characteristic energies are combined in Figs.~\ref{fig2}(c) and \ref{fig2}(d) for spectra taken with Pt/Ir-tips over two different $(500 \times 500) \rm nm^2$ areas, and in Figs.~\ref{fig2}(g) and \ref{fig2}(h) for spectra taken with Cr-coated tips over two different $(500 \times 500) \rm nm^2$ areas. In general, we find that $\Delta _{\pm} \sim 0.6$ eV and $\Delta _{\pm} ^{\ast} \sim 0.4$ eV at 77 K for data taken with Pt/Ir tips. The apparent spatial variations in the characteristic energies are manifestations of the intrinsically heterogeneous nature of the manganites, even for the most conducting and homogeneous LCMO composition considered in this work.

While the characteristics of each type of spectra taken with Pt/Ir and Cr-coated tips were qualitatively similar, as exemplified in Figs.~\ref{fig2}(a) and \ref{fig2}(e) and also manifested by the histograms shown in Figs.~\ref{fig2}(c) and \ref{fig2}(g) and Figs.~\ref{fig2}(d) and \ref{fig2}(h), careful inspections of the spectral details reveal quantitative differences. These differences may be attributed to the energy-dependent DOS of Cr, which will be discussed further in Section IV. Therefore, in the following discussion we only refer to the data taken with Pt/Ir tips as representative of the DOS of LCMO. 

According to bandstructure calculations,~\cite{Pickett96,Satpathy96} the values of $U_{\pm}$ in the bulk DOS of LCMO are correlated directly with the Ca-doping level and are well defined for a given Ca-doping level $x$, as exemplified in Fig.~\ref{fig3}(a) for specific doping levels considered in bandstructure calculations. Therefore, the finite range of $U_{\pm}$ values manifested by the histograms in Figs.~\ref{fig2}(c), \ref{fig2}(d), \ref{fig2}(g) and \ref{fig2}(h) imply spatially varying Ca-doping levels. Nonetheless, the dominating value of $|U_{\pm}|$ varies between 1.0 and 1.2 eV, which correspond to local Ca-doping levels between $x = 0.33$ and $x = 0.25$, in good agreement with the nominal doping level of our sample $x = 0.3$ if we estimate the Ca-doping level by assuming monotonic $U_{\pm}$-vs.-$x$ dependence as shown in Fig.~\ref{fig3}(a). Additionally, the apparent positive correlation between most regions of the maps of $U_+$ and $\Delta _+$ values over the same ($500 \times 500$) $\rm nm^2$ sample area, as shown in Figs.~\ref{fig3}(b) -- \ref{fig3}(c) and further quantified by the map of cross correlations~\cite{correlation} between $U_+$ and $\Delta _+$ in Fig.~\ref{fig3}(d), is suggestive of a common physical origin for the spatial variation in $U_+$ and $\Delta _+$. In other words, the intrinsic electronic heterogeneity in the manganites is responsible for the empirical observation of spectral variations at low temperatures and in zero-field. Further, from Fig.~\ref{fig3}(a) we find that positive correlation between $U_+$ and $\Delta _+$ implies that larger $\Delta _+$ values are associated with lower Ca-doping levels (larger $U_+$ values).

\begin{figure*}
  \centering
  \includegraphics[width=5.2in]{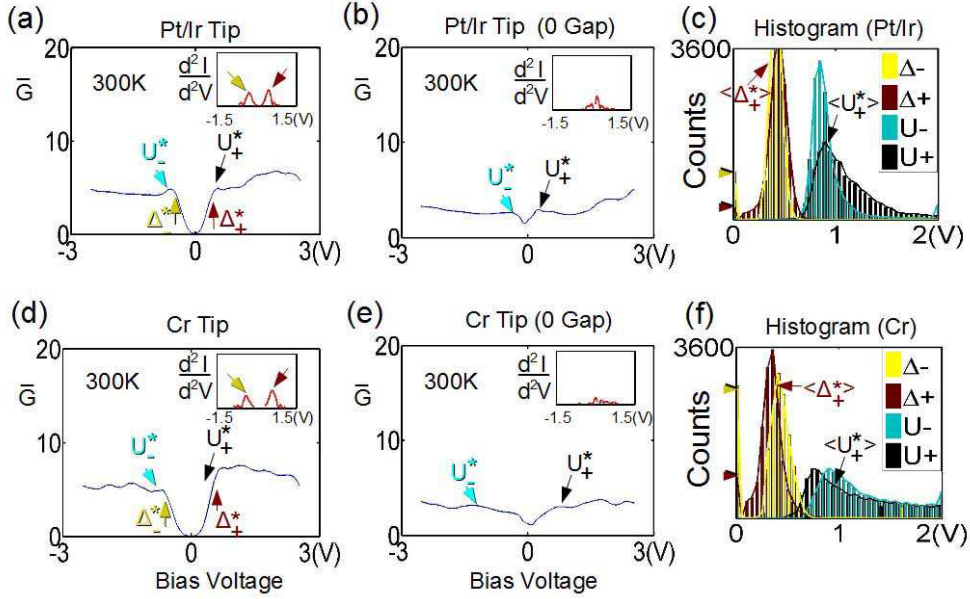}
\caption{(Color online) Comparison of the tunneling spectral characteristics taken with Pt/Ir and Cr-coated tips at $T = 300$ K and $H = 0$: (a) A PG-like spectrum taken with a Pt/Ir tip, showing significantly suppressed $U_{\pm} ^{\ast}$ values relative to the peak energies $U_{\pm}$ found in the spectrum of Fig.~2(a). The PG values $\Delta _{\pm} ^{\ast}$ are determined from the peaks of the $(d^2I/dV^2)$-vs.-$V$ spectrum. (b) Another typical type of spectra taken with a Pt/Ir tip, showing vanishing gaps as detailed in the inset. (c) Histograms of the PG values $\Delta _{\pm} ^{\ast}$ and the characteristic energies $U_{\pm} ^{\ast}$ obtained by using a Pt/Ir tip over a $(500 \times 500) \rm nm^2$ area at 300 K, showing suppressed $U_{\pm} ^{\ast}$ values relative to the $U_{\pm}$ values found at 77 K, as well as large counts of vanishing gaps (shown by the arrows at $\omega = 0$) and PG at $\omega = \langle \Delta _{\pm} ^{\ast} \rangle$. (d) A typical PG-like spectrum taken with a Cr-coated tip, showing suppressed $U_{\pm} ^{\ast}$ values relative to the peak energies $U_{\pm}$ found in the spectrum of Fig.~2(e). The PG values $\Delta _{\pm} ^{\ast}$ are determined from the peaks of the $(d^2I/dV^2)$-vs.-$V$ spectrum. (e) Another typical type of spectra taken with a Cr tip, showing vanishing gaps as detailed in the inset. (f) Histograms of PG values $\Delta _{\pm} ^{\ast}$ and the characteristic energies $U_{\pm} ^{\ast}$ obtained by using a Cr-coated tip over a $(500 \times 500) \rm nm^2$ area at 300 K, showing suppressed $U_{\pm} ^{\ast}$ values relative to the $U_{\pm}$ values found at 77 K, as well as large counts of vanishing gaps (shown by the arrows at $\omega = 0$) and PG at $\omega = \langle \Delta _{\pm} ^{\ast} \rangle$.}
\label{fig4}
\end{figure*}

As noted before, the sample areas studied with different STM tips and at different temperatures are generally not identical. Therefore, for meaningful comparison of the spectral evolution with temperature and spin polarization, it is necessary to establish statistical consistency of the spectral characteristics obtained from one $(500 \times 500) \ \rm nm^2$ area with those obtained from another $(500 \times 500) \rm \ nm^2$ area of the same sample. Similarly, for different samples prepared under the same fabrication conditions, it is also necessary to establish the same statistical consistency. In this context, we show in Figs.~2(c), 2(d), 2(g) and 2(h) the histograms of $U_{\pm}$ and $\Delta _{\pm}$ obtained from the tunneling spectra on different sample areas at 77 K with Pt/Ir and Cr-coated tips. As evidenced by the similarities of Fig.~2(c) to Fig.~2(d) and Fig.~2(g) to Fig.~2(h), the spectral characteristics over different $(500 \times 500) \ \rm nm^2$ areas appear to be statistically consistent for the same type of STM tips and at the same temperature.  

\begin{figure}
  \centering
  \includegraphics[width=3.4in]{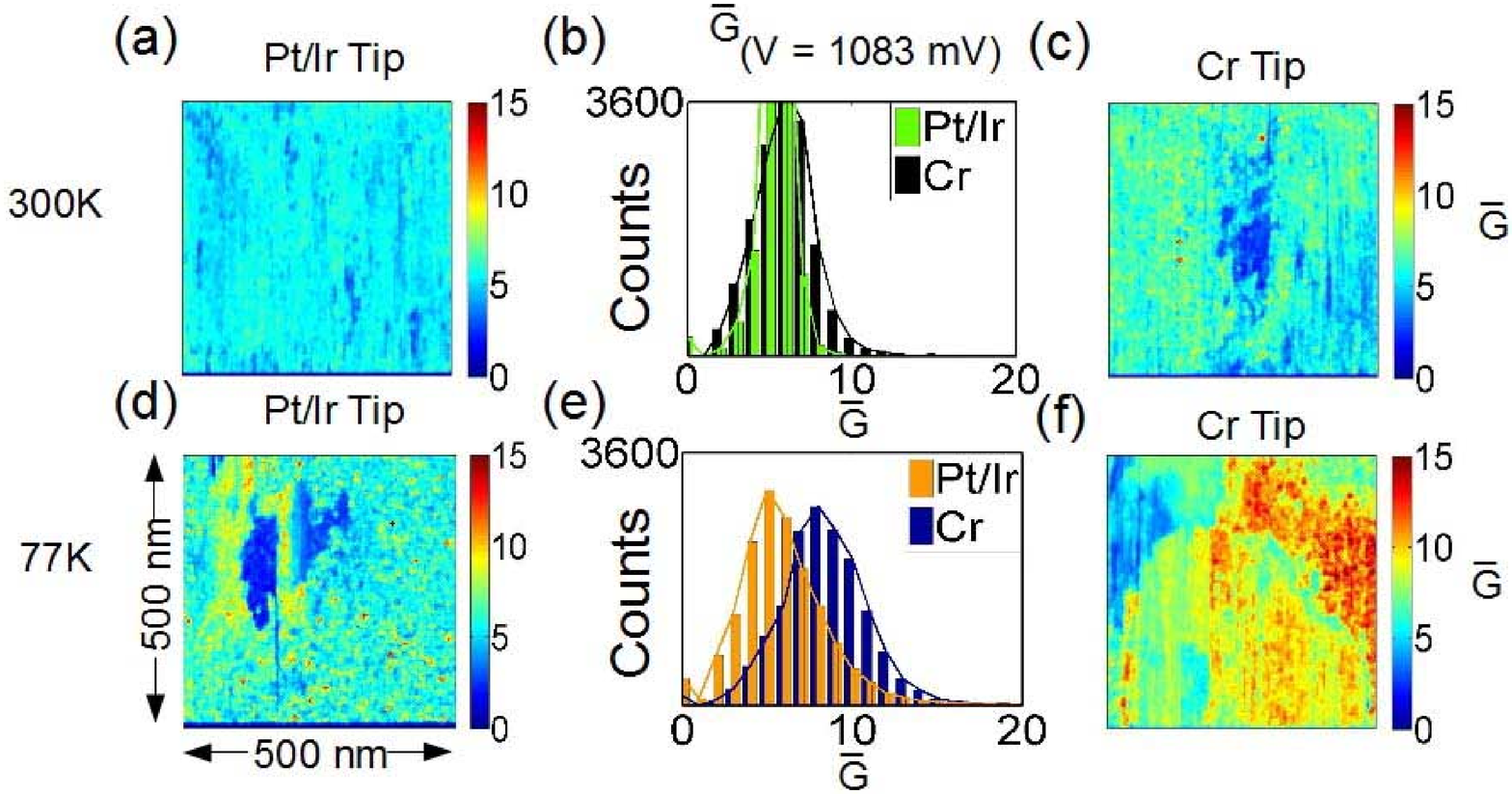}
\caption{(Color online) Comparison of the high-bias tunneling spectral characteristics taken with Pt/Ir and Cr-coated tips at $H = 0$: (a) A $(500 \times 500) \rm nm^2$ tunneling conductance map taken with a Pt/Ir tip at $\omega = \langle U_+ ^{\ast} \rangle$ and $T = 300$ K. (b) Histograms of the tunneling conductance obtained by using a Pt/Ir tip and a Cr-coated tip at $T = 300$ K and for $\omega = \langle U_+ ^{\ast} \rangle$. (c) A $(500 \times 500) \rm nm^2$ tunneling conductance map taken with a Cr-coated tip at $\omega = \langle U_+ ^{\ast} \rangle$ and $T = 300$ K. (d) A $(500 \times 500) \rm nm^2$ tunneling conductance map taken with a Pt/Ir tip at $\omega = \langle U_+ \rangle$ and $T = 77$ K. (e) Histograms of the tunneling conductance obtained by using a Pt/Ir tip and a Cr-coated tip at $T = 77$ K and for $\omega = \langle U_+ \rangle$.(f) A $(500 \times 500) \rm nm^2$ tunneling conductance map taken with a Cr-coated tip at the characteristic energy $\omega = \langle U_+ \rangle$ and $T = 77$ K. }
\label{fig5}
\end{figure}

\subsection{Temperature dependence}

\begin{figure}
  \centering
  \includegraphics[width=3.4in]{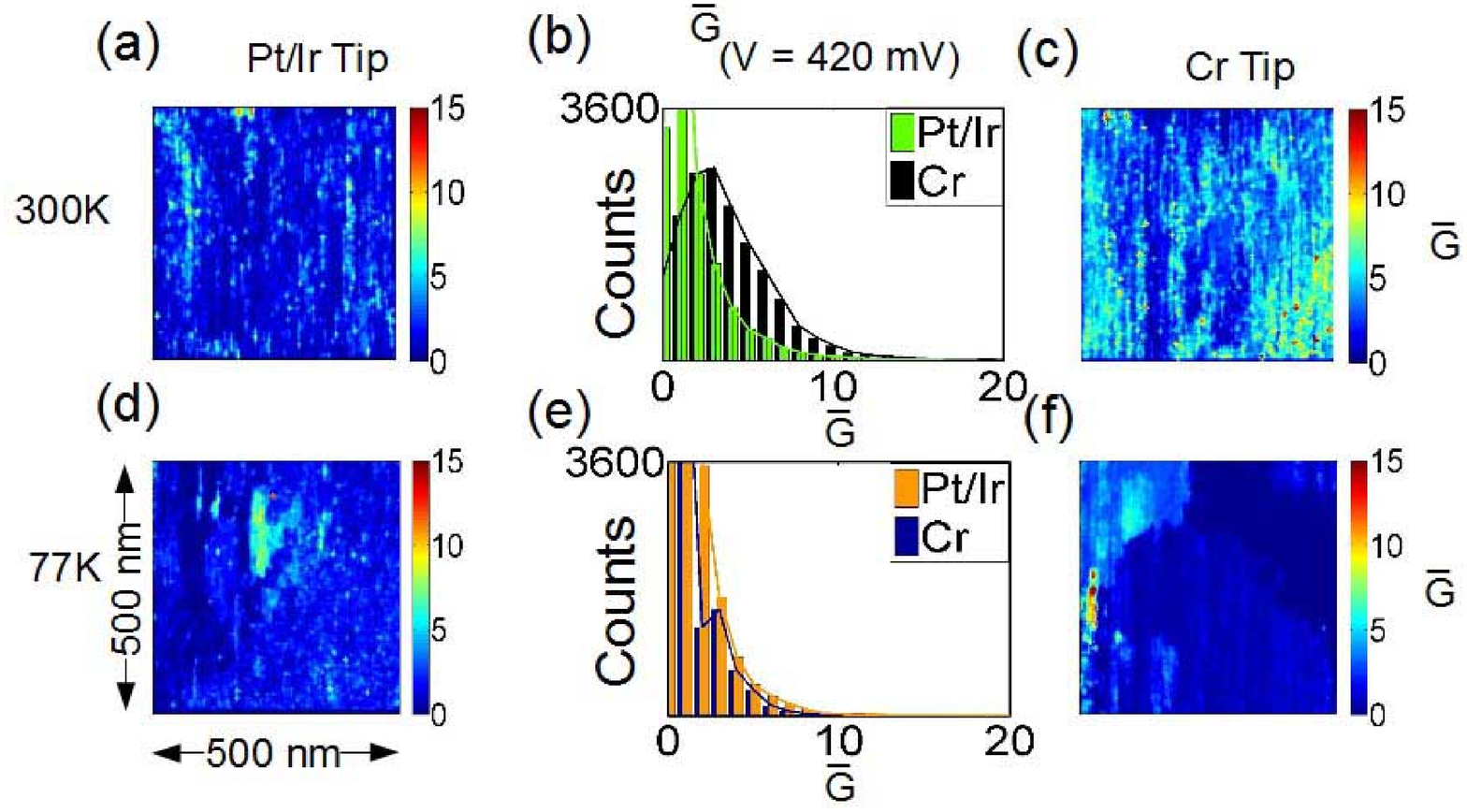}
\caption{(Color online) Comparison of the low-bias tunneling spectral characteristics taken with Pt/Ir and Cr-coated tips at $H = 0$: (a) A $(500 \times 500) \rm nm^2$ tunneling conductance map taken with a Pt/Ir tip at $\omega = \langle \Delta _+ ^{\ast} \rangle$ and 300 K. (b) Histograms of the tunneling conductance obtained by using a Pt/Ir tip and a Cr-coated tip at 300 K for $\omega = \langle \Delta _+ ^{\ast} \rangle$. (c) A $(500 \times 500) \rm nm^2$ tunneling conductance map taken with a Cr-coated tip at $\omega = \langle \Delta _+ ^{\ast} \rangle$ and for $T$ = 300 K. (d) A $(500 \times 500) \rm nm^2$ tunneling conductance map taken with a Pt/Ir tip at $\omega = \langle \Delta _+ \rangle$ and $T$ = 77 K. (e) Histograms of the tunneling conductance obtained by using a Pt/Ir tip and a Cr-coated tip at 77 K and for $\omega = \langle \Delta _+ \rangle$. (f) A $(500 \times 500) \rm nm^2$ tunneling conductance map taken with a Cr-coated tip at the characteristic energy $\omega = \langle \Delta _+ \rangle$ and for $T$ = 77 K.}
\label{fig6}
\end{figure}

At $T = 300$ K the LCMO epitaxial thin films studied in this work were in the paramagnetic phase. Hence, the spectral characteristics at $T = 300$ K were quite different from those observed in the bulk ferromagnetic state at $T = 77$ K, as shown in Figs.~\ref{fig4}(a)-(b) and \ref{fig4}(d)-(e) for exemplified spectra taken at 300 K and with Pt/Ir and Cr-coated tips, respectively. Comparison of Figs.~\ref{fig4}(a)-(b) and \ref{fig4}(d)-(e) with Figs.~2(a)-(b) and 2(e)-(f) indicates several important contrasts. First, the large DOS peaks associated with $U_{\pm}$ for ferromagnetic LCMO became much suppressed in the paramagnetic state and the $U_{\pm}^{\ast}$ values become much smaller than $U_{\pm}$. Second, the surface insulating gap found around the Fermi level at 77 K either completely disappeared (Figs.~\ref{fig4}(b) and \ref{fig4}(e)) or became a PG (Figs.~4(a) and 4(d)), as summarized by the histograms of the PG $\Delta _{\pm} ^{\ast}$ in Figs.~\ref{fig4}(c) and \ref{fig4}(f), where large counts at both zero and the PG energies $\langle \Delta _+ ^{\ast} \rangle$ are shown. The vanishing gaps at 300 K for some of the spectra cannot be accounted for by thermal smearing alone, and are therefore suggestive of a magnetic phase transition occurring at a mean transition temperature between 77 K and 300 K. On the other hand, the nearly temperature independent PG energies are suggestive of a completely different physical origin. Third, slight differences were found between the spectra taken with the Pt/Ir tip and those taken with the Cr-coated tip at 300 K, as manifested in Figs.~\ref{fig4}(a)-(b) and \ref{fig4}(d)-(e). The differences occurred because spectra taken with the former were representative of the DOS of LCMO in the paramagnetic phase, whereas those taken with the latter consisted of convoluted DOS of the paramagnetic LCMO and the ferromagnetic Cr-coated tip. 

\begin{figure*}
  \centering
  \includegraphics[width=5.2in]{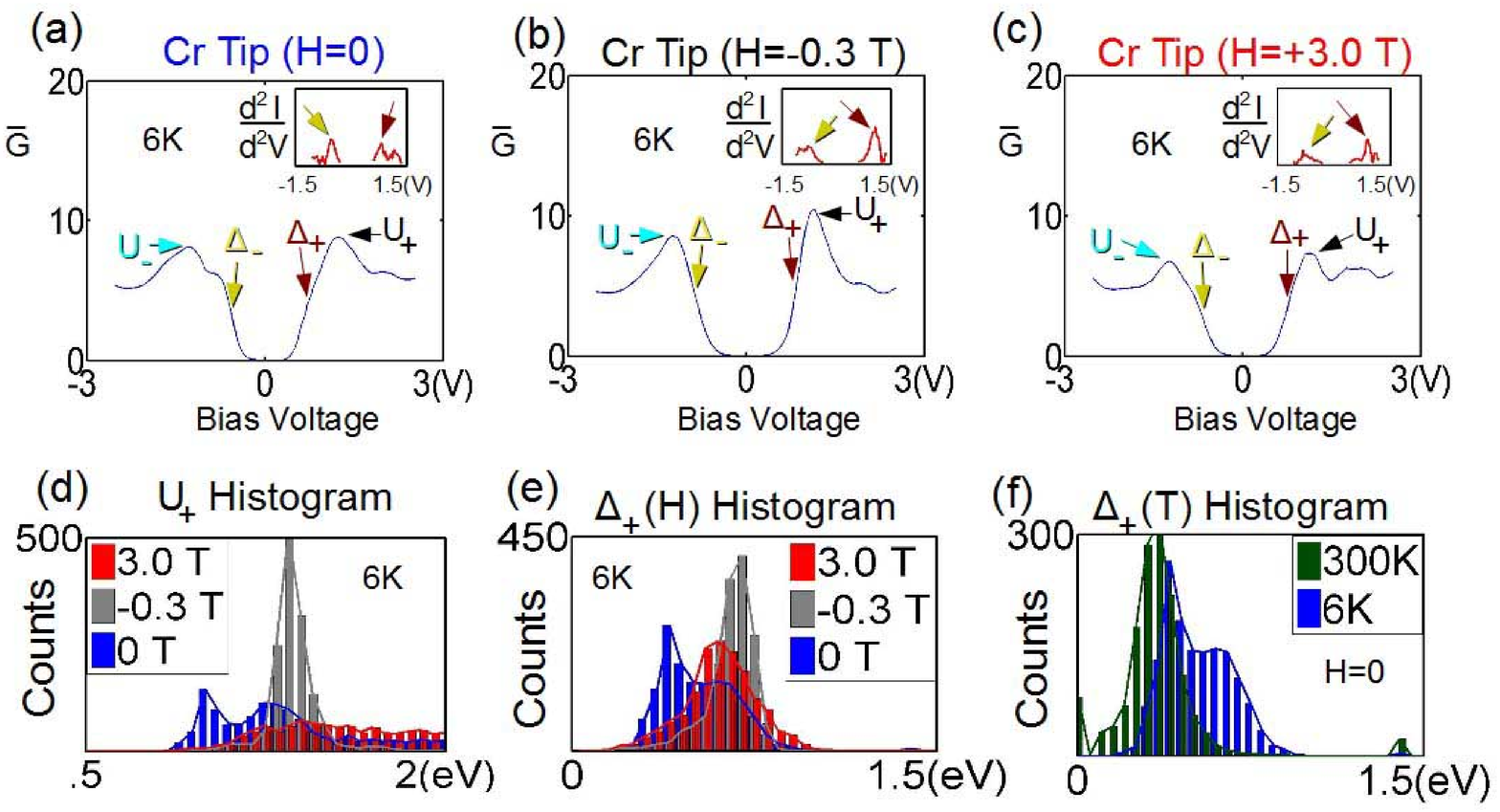}
\caption{(Color online) Comparison of magnetic field-dependent spectral characteristics taken with a Cr-coated tip at $T = 6$ K: (a) Normalized conductance $(dI/dV)/(I/V) \equiv \bar{G}$ vs. bias voltage ($V$) spectrum for $H = 0$. (b) $\bar{G}$ vs. $V$ spectrum for $H = -0.3$ T. (c) $\bar{G}$ vs. $V$ spectrum for $H = 3.0$ T. (d) Histograms of the characteristic energy $U_+$ over the same $(250 \times 90) \rm nm^2$ sample area for $H = 0$, $-0.3$ T and 3.0 T. (e) Histograms of $\Delta_+$ over the same $(250 \times 90) \rm nm^2$ sample area at $H = 0$, $-0.3$ T and 3.0 T. (f) Temperature evolution of the histograms of $\Delta_+$ over a $(250 \times 90) \rm nm^2$ sample area at $H = 0$, showing downshifts in the insulating gap values with increasing temperature. Specifically, two types of gap values at 6 K may be attributed to an insulating surface gap and a pseudogap. The former vanishes and the latter persists at 300 K.}
\label{fig7}
\end{figure*}

In addition to the temperature-dependent spectral characteristics, the spatial variations in the tunneling conductance also revealed temperature-dependent evolution. Specifically, the tunneling conductance in the paramagnetic state was generally more homogeneous than that in the ferromagnetic state, because the tendency toward phase separations only occurred in the ferromagnetic state of LCMO, as manifested by the constant-bias tunneling conductance maps in Figs.~\ref{fig5}(a) and \ref{fig5}(c) for room temperature spectra taken at $\omega = \langle U_+ ^{\ast }\rangle$ with the Pt/Ir and Cr-coated tips, respectively. In contrast, the tunneling conductance in the ferromagnetic state was significantly more inhomogeneous, as exemplified in Figs.~5(d) and 5(f) for tunneling conductance taken at 77 K and for $\omega = \langle U_+ \rangle$. Here $\langle U_+ \rangle$ is defined as the most commonly occurring $U_+$ values obtained from the histograms in Fig.~2. The statistical distributions of the conductance at $\omega = \langle U_+ \rangle$ for 77 K and at $\omega = \langle U_+ ^{\ast} \rangle$ for 300 K are summarized by the histograms in Figs.~\ref{fig5}(b) and \ref{fig5}(e) for spectra taken with the Pt/Ir and Cr-coated tips, respectively. While the histograms at $T$ = 300 K were statistically similar between spectra taken with Pt/Ir and Cr-coated tips as shown in Fig.~\ref{fig5}(b), at 77 K the tunneling conductance distributions for spectra taken with the Cr-coated tip revealed an overall shift towards higher conductance than those taken with the Pt/Ir tip. The apparent differences between the histograms obtained with Pt/Ir and Cr-coated tips from LCMO at 77 K are suggestive of different effects associated with regular and spin-polarized tunneling into spatially inhomogeneous LCMO in its ferromagnetic phase. 

\begin{figure}
  \centering
  \includegraphics[width=2.95in]{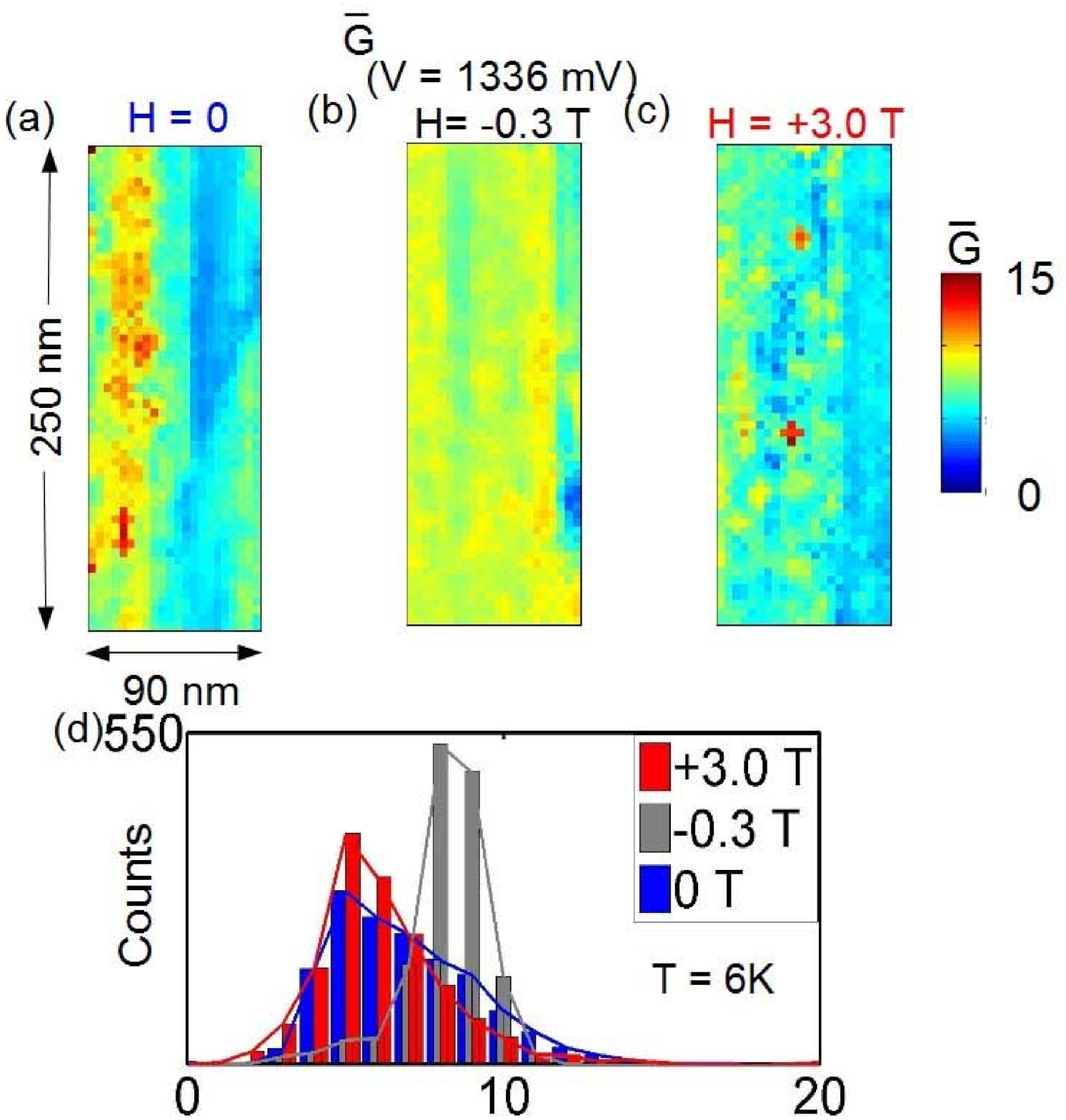}
\caption{(Color online) Comparison of magnetic field-dependent high-bias conductance maps ($\omega = \langle U_+ \rangle$) taken over the same $(250 \times 90) \rm nm^2$ sample area with a Cr-coated tip at $T = 6$ K: (a) Normalized conductance map for $H = 0$. (b) Normalized conductance map for $H = -0.3$ T. (c) Normalized conductance map for $H = 3.0$ T. (d) Histograms of the normalized conductance at $\omega = \langle U_+ \rangle$ and for $H = 0$, $-0.3$ T and 3.0 T, showing highest mean conductance at $H = -0.3$ T when the spin-polarization of the tunneling currents is antiparallel to the magnetization of LCMO.}
\label{fig8}
\end{figure}

Similarly, the tunneling conductance maps for $\omega = \langle \Delta _+ ^{\ast} \rangle$ taken at 300 K with Pt/Ir and Cr-coated tips are shown in Figs.~\ref{fig6}(a) and \ref{fig6}(c), respectively, whereas those for $\omega = \langle \Delta _+ \rangle$ taken at 77 K with Pt/Ir and Cr-coated tips are shown in Figs.~\ref{fig6}(d) and \ref{fig6}(f). These maps again reveal spatially more homogeneous tunneling conductance in the paramagnetic state. For completeness, the statistical distributions of the tunneling conductance at $\omega = \langle \Delta_+ ^{\ast} \rangle$ for $T = 300$ K and $\omega = \langle \Delta_+ \rangle$ for $T = 77$ K are summarized by the histograms in Figs.~\ref{fig6}(b) and \ref{fig6}(e). Here $\langle \Delta _+ \rangle$ denotes the most commonly occurring insulating gap value at positive bias from the histograms in Fig.~2 for $T = 77$ K, and $\langle \Delta _+ ^{\ast} \rangle$ represents the most commonly found PG values from the histograms in Fig.~4 for $T = 300$ K.  

\subsection{Magnetic field dependence}

\begin{figure}
  \centering
  \includegraphics[width=3.0in]{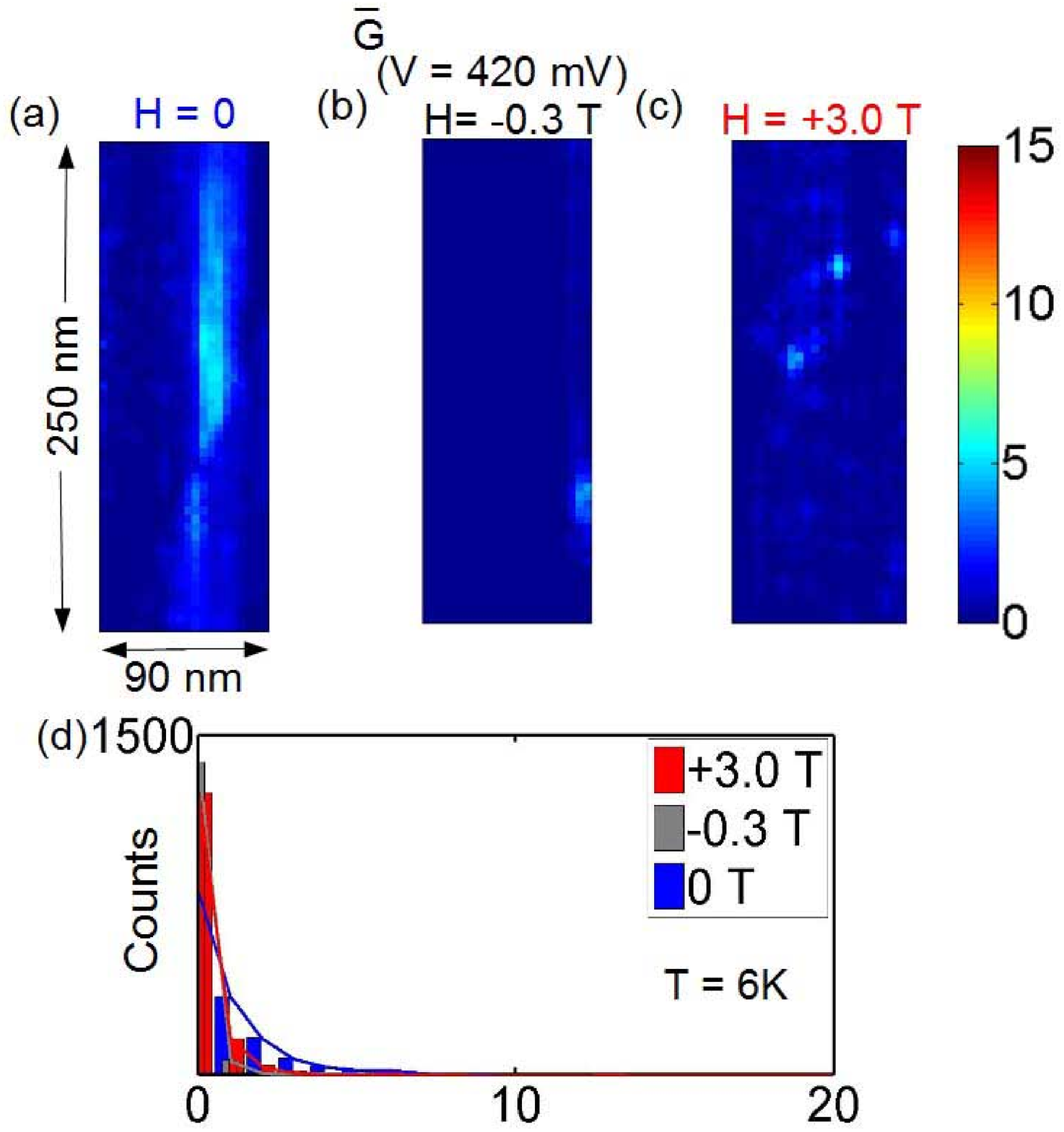}
\caption{(Color online) Comparison of magnetic field-dependent low-bias conductance maps ($\omega = \langle \Delta _+ \rangle$) taken over the same $(250 \times 90) \rm nm^2$ sample area with a Cr-coated tip at $T = 6$ K: (a) Normalized conductance map for $H = 0$. (b) Normalized conductance map for $H = -0.3$ T. (c) Normalized conductance map for $H = 3.0$ T. (d) Histograms of the normalized conductance at $\langle \Delta _+ \rangle$ at $H = 0$, $-0.3$ T and 3.0 T, showing lowest mean conductance at $H = -0.3$ T when the spin-polarization of the tunneling currents is antiparallel to the magnetization of LCMO.}
\label{fig9}
\end{figure}

Although differences between the tunneling spectra taken with Pt/Ir tips and those taken with Cr-coated tips are readily visible at $H = 0$ and for $T < T_C ^{\rm LCMO}$, additional magnetic field dependent investigations are necessary to provide better quantitative understanding for spin-polarized tunneling in LCMO. As described previously, the degree of spin polarization may be controlled by keeping $T < T_C ^{\rm LCMO}$ and by applying oppositely directed magnetic fields with magnitudes satisfying either the condition $H_C^{\rm LCMO} < |H| < H_C^{\rm Cr}$ or $|H| > H_C^{\rm Cr}$. The applied fields $H = -0.3$ T and $H = 3.0$ T chosen in this work are consistent with the required conditions. 

\begin{figure*}
  \centering
  \includegraphics[width=5.2in]{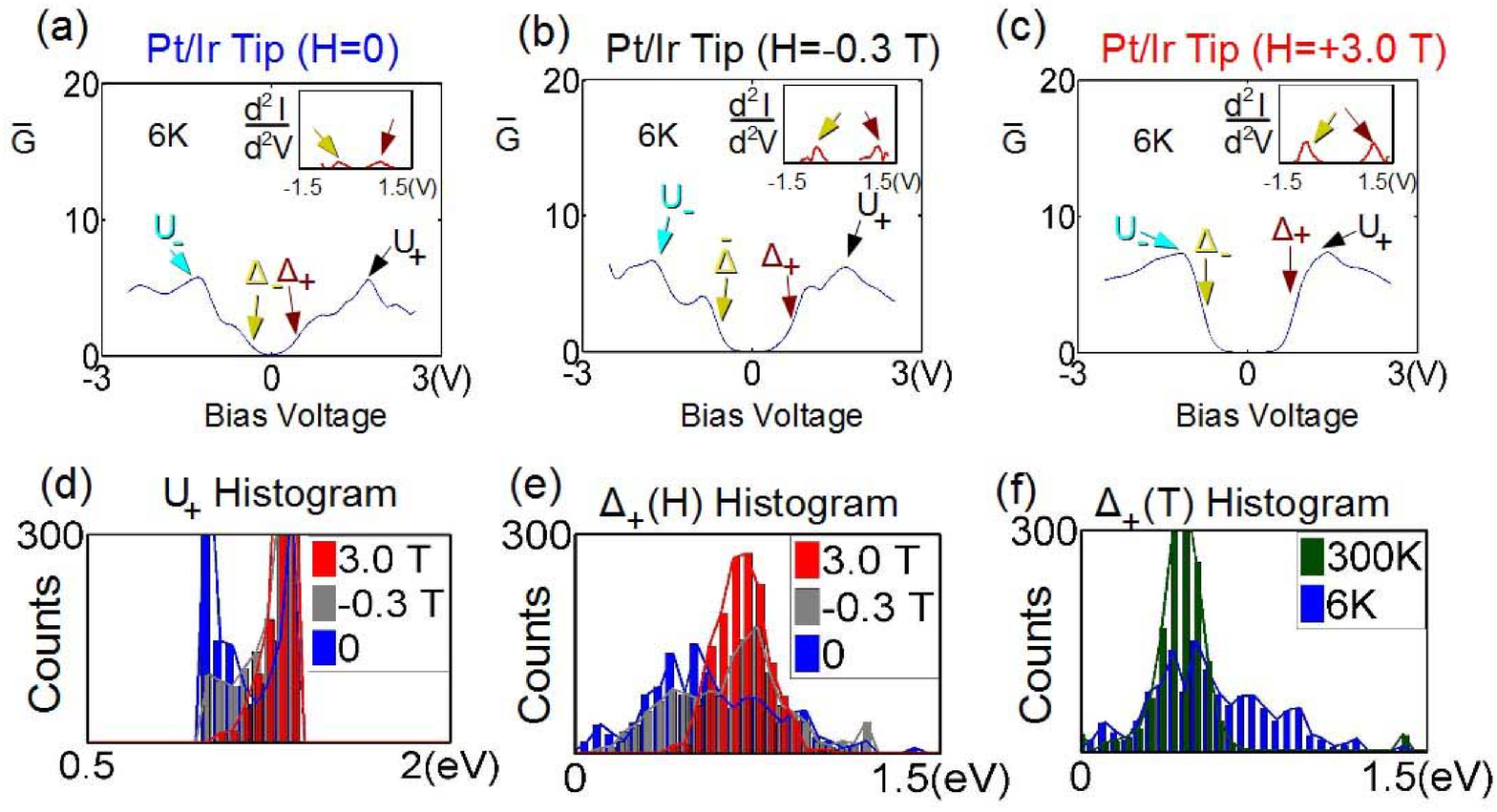}
\caption{(Color online) Comparison of magnetic field-dependent spectral characteristics taken with a Pt/Ir tip at $T = 6$ K: (a) Normalized conductance $\bar{G} \equiv (dI/dV)/(I/V)$ vs. energy ($\omega$) spectrum for (a) $H = 0$, (b) $H = -0.3$ T and (c) $H = 3.0$ T. (d) Histograms of the characteristic energies $U_+$ obtained from the same $(250 \times 28) \rm nm^2$ sample area under $H = 0$, $-0.3$ T and 3.0 T, showing monotonic shift towards the right with increasing $|H|$. (e) Histograms of surface gap energies $\Delta_+$ obtained from the same $(250 \times 28) \rm nm^2$ sample area under $H = 0$, $-0.3$ T and 3.0 T, showing monotonic increase of $\langle \Delta _+ \rangle$ with increasing $|H|$. (f) Comparison of the zero-field temperature evolution of $\Delta _+$, showing a wide distribution of insulating gap values at 6 K and a persistent pseudogap at $\Delta _+ ^{\ast} \sim 0.4$ eV at 300 K.}
\label{fig10}
\end{figure*}

In Figs.~\ref{fig7}(a)-(c) representative normalized spectra taken in the same area with a Cr-coated tip and at $T = 6$ K are shown for $H =0$, $-0.3$ T and $3.0$ T. It is apparent that these tunneling spectra evolved significantly with magnetic field, with the statistical field-dependent spectral evolution for a $(250 \times 90) \rm nm^2$ sample area summarized by the histograms of the characteristic energies $U_+$ and $\Delta _+$ in Figs.~\ref{fig7}(d)-(e). In particular, the non-monotonic field dependence of $\langle U_+ \rangle$ and $\langle \Delta _+ \rangle$ is noteworthy. Further, the normalized tunneling conductance map also revealed significant and non-monotonic field dependence, as exemplified in Figs.~\ref{fig8}(a)-(c) and in Figs.~\ref{fig9}(a)-(c) for spatially resolved tunneling conductance maps taken at the characteristic energies $\omega = \langle U_+ \rangle$ and $\langle \Delta _+ \rangle$, respectively, and for $H =0$, $-0.3$ T and $3.0$ T over the same $(250 \times 90) \rm nm^2$ sample area. We find that the spatially inhomogeneous conductance map at $H = 0$ and for $\omega = \langle U_+ \rangle$ became more homogeneous in finite fields, reaching overall highest conductance for $H = -0.3$ T, as manifested statistically by the histograms of conductance in Figs.~\ref{fig8}(a)-(c) for $H =0$, $-0.3$ T and $3.0$ T. Additionally, all conductance maps taken at $\omega = \langle U_+ \rangle$ appear to correlate with those at $\omega = \langle \Delta _+ \rangle$ when we compare Figs.~\ref{fig8}(a)-(c) with Figs.~\ref{fig9}(a)-(c).

\begin{figure}
  \centering
  \includegraphics[width=3.1in]{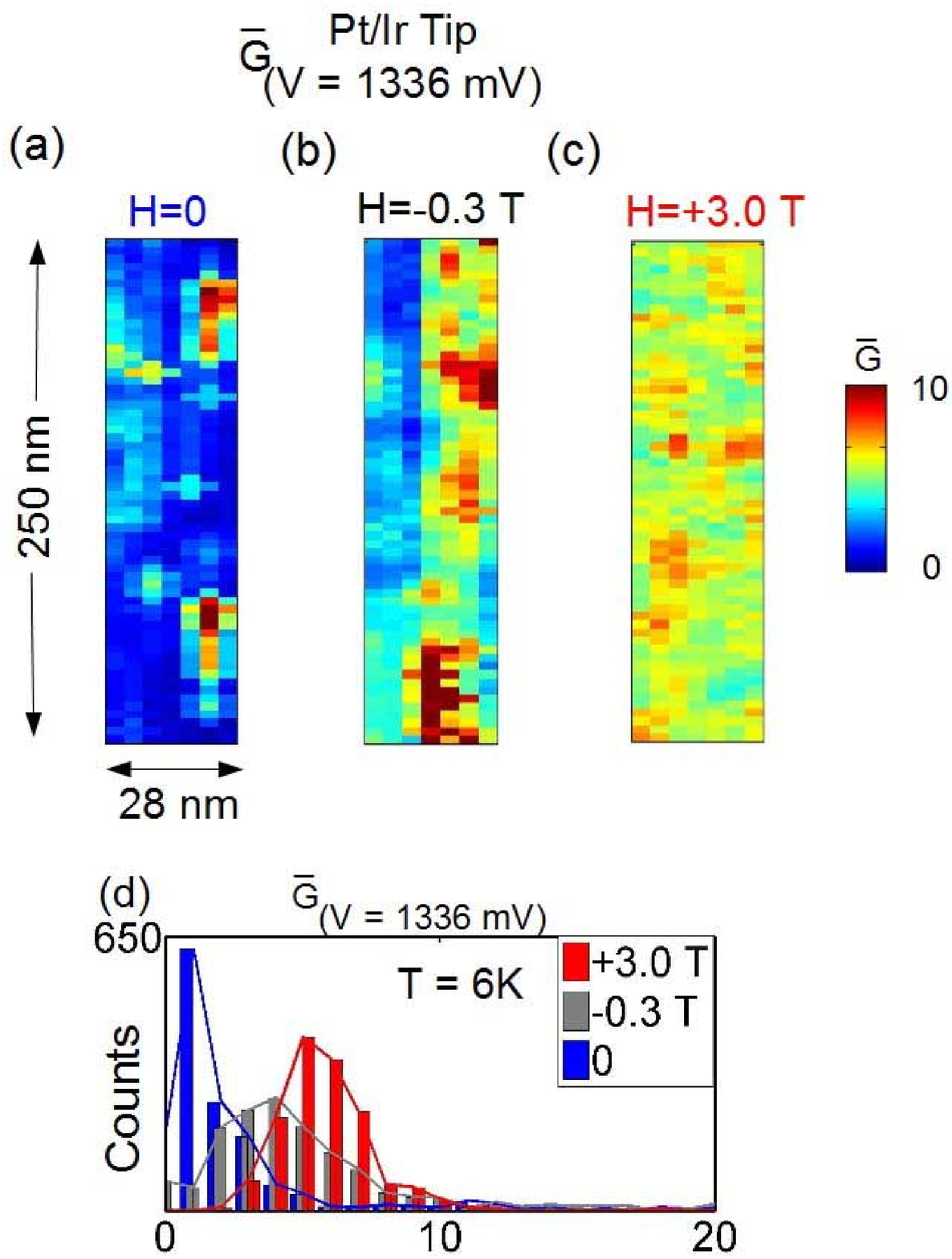}
\caption{(Color online) Comparison of magnetic field-dependent high-bias tunneling conductance maps taken at $\omega = \langle U_+ \rangle \sim 1336$ meV over the same $(250 \times 28)$ sample area with a Pt/Ir tip at $T = 6$ K and for (a) $H = 0$, (b) $H = -0.3$ T, and (c) $H = 3.0$ T, showing monotonic increase in the homogeneity and the value of $\bar{G}$ with increasing $|H|$. (d) Histograms of the normalized tunneling conductance $\bar{G}$ at $\omega = \langle U_+ \rangle$ at $H = 0$, $-0.3$ T and 3.0 T.}
\label{fig11}
\end{figure}

\begin{figure}
  \centering
  \includegraphics[width=3.1in]{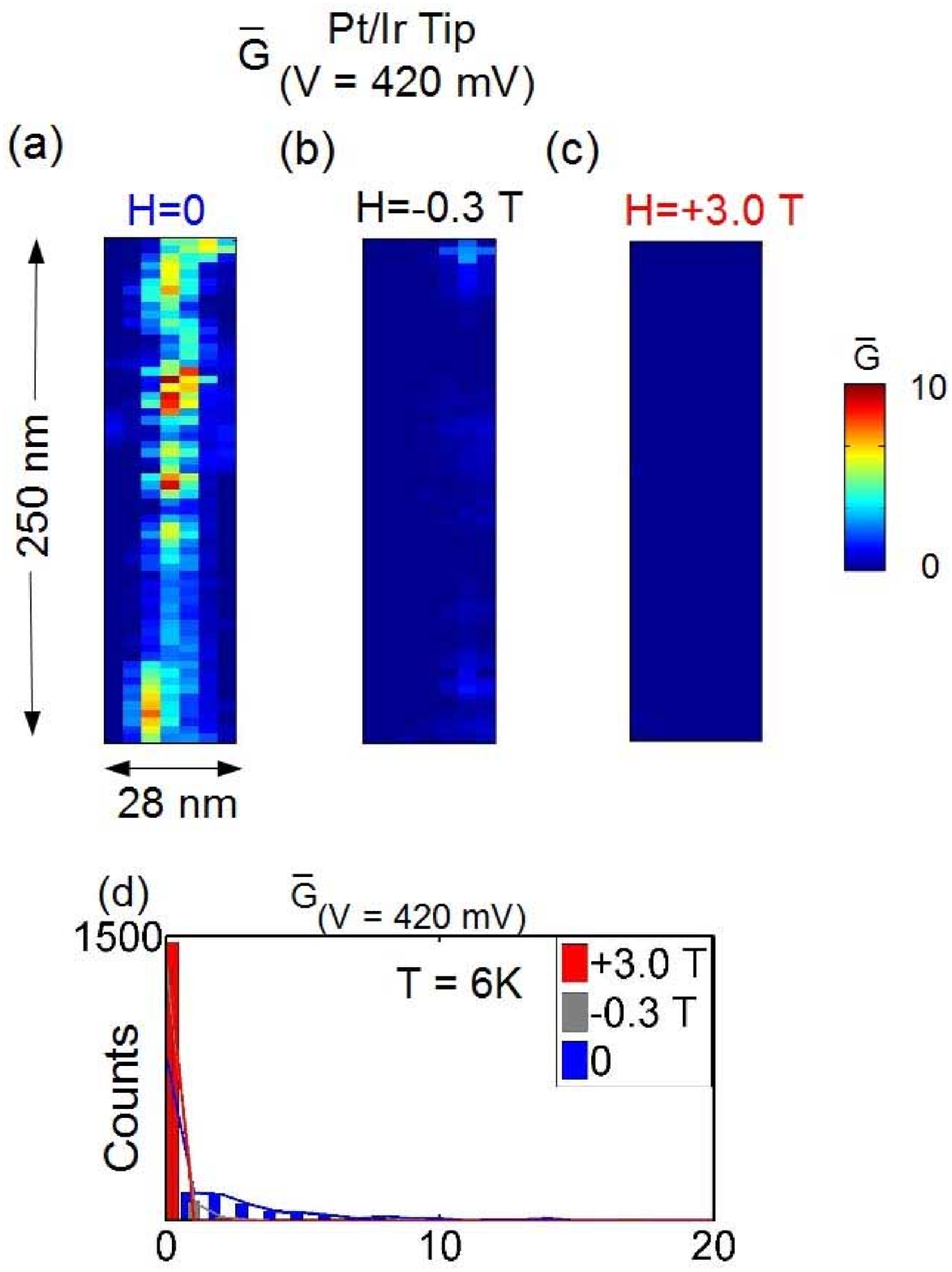}
\caption{(Color online) Comparison of magnetic field-dependent low-bias tunneling conductance maps taken at $\omega = \langle \Delta _+ \rangle \sim 420$ meV over the same sample area as in Fig.~11 with a Pt/Ir tip at $T = 6$ K and for (a) $H = 0$, (b) $H = -0.3$ T and (c) $H = 3.0$ T, showing monotonic increase in the conductance homogeneity but decrease in the value of $\bar{G}$ with increasing $|H|$. (d) Histograms of $\bar{G}$ at $\omega = \langle \Delta _+ \rangle$ and for $H = 0$, $-0.3$ T, and 3.0 T.}
\label{fig12}
\end{figure}

For comparison, similar magnetic field-dependent spectroscopic studies were conducted under the same conditions with a Pt/Ir tip. The representative tunneling spectra taken in the same area with a Pt/Ir tip at $T = 6$ K and for $H = 0$, $-0.3$ T and $3.0$ T are shown in Figs.~10(a)-(c). Overall these tunneling spectra exhibit different field-dependent evolution when compared with the spectra taken with a Cr-coated tip, as statistical manifested by the histograms of the characteristic energies $U_+$ and $\Delta _+$ in Figs.~10(d)-(e). In particular, we note that the histograms of $U_+$ values appear to shift monotonically up to higher energies with the increasing magnitude of magnetic field. Similarly, the insulating gap values $\Delta _+$ appear to shift up monotonically and the distributions become sharper with the magnitude of increasing magnetic field. The spectral characteristics obtained with Pt/Ir tips are therefore only dependent on the magnitude of applied fields, which are in stark contrast to the spectra obtained with Cr-coated tips that are strongly dependent on the direction of the applied magnetic field. Additionally, the tunneling conductance $\bar{G}$ taken with Pt/Ir tips at $\omega = \langle U_+ \rangle$ generally increases with increasing $|H|$, as exemplified by the conductance maps in Figs.~11(a)-(c) for $H = 0$, $-0.3$ T and $3.0$ T and also summarized by the conductance histograms in Fig.~11(d). 

The general trend of increasing tunneling conductance and spatial homogeneity at $\omega = \langle U_+ \rangle$ with increasing magnetic field for data taken with the Pt/Ir tip is consistent with the CMR nature of the manganites, because better alignment of magnetic domains and increasing mobility with increasing magnetic field results in enhanced electrical conductance across different magnetic domains. We further note that the increasing spatial homogeneity in the tunneling conductance with increasing $|H|$ also agrees with previous STS reports on LCMO,~\cite{Fath99} although previous reports focused on tunneling conductance studies near the Curie temperature and only at one constant energy $\omega = 3.0$ eV. Additionally, we find that the conductance $\bar{G}$ at $\omega = \langle \Delta _+ \rangle$ {\it decreases} with increasing $|H|$, as shown in Figs.~12(a)-(c). The opposite field-dependence of $\bar{G} (\omega)$ for $\omega = \langle \Delta _+ \rangle$ to that for $\omega = \langle U_+ \rangle$ suggests that $U_+$ and $\Delta _+$ are associated with different characteristics of the LCMO sample. This point will be elaborated in the following analysis and discussion.  

\begin{figure}
  \centering
  \includegraphics[width=3.3in]{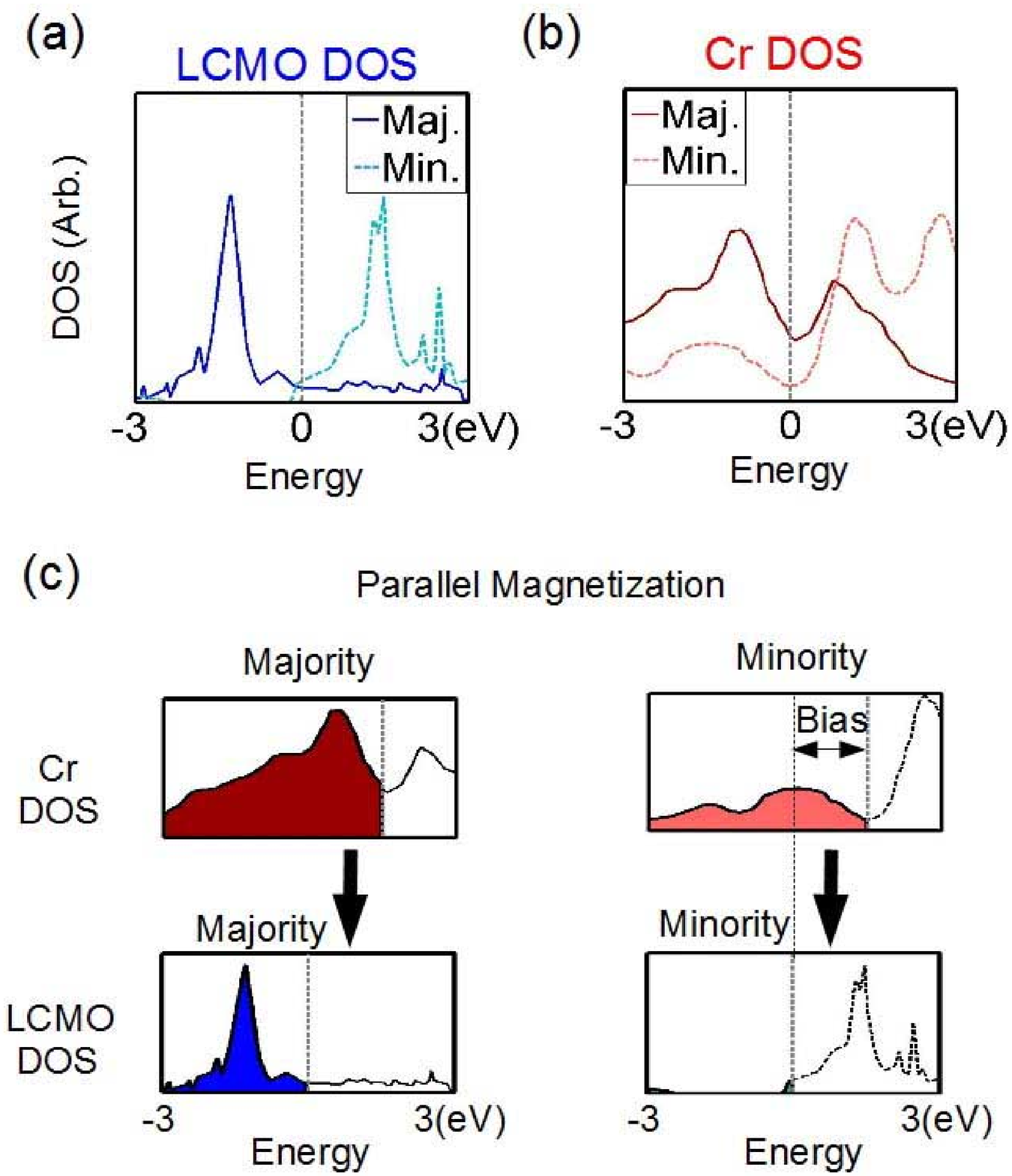}
\caption{(Color online) Theoretical density of states (DOS) for LCMO and Cr at $T \ll T_C$ from bandstructure calculations: (a) Theoretical DOS for the majority and minority bands of LCMO with $x = 0.25$.~\cite{Pickett96} (b) Theoretical DOS for the majority and minority bands of the first layer of Cr in a thin film. (c) Theoretical tunneling configurations for parallel magnetization between Cr-coated tip and LCMO sample under a positive bias voltage $V = (U_+/e)$, showing the dominating processes of electron tunneling from the occupied majority band of Cr to the empty majority band of LCMO and from the occupied minority band of Cr to the empty minority band of LCMO.}
\label{fig13}
\end{figure}

The apparent contrasts between aforementioned field-dependent spectra taken at 6 K with Cr-coated tips and those taken with Pt/Ir tips, as shown in Figs.~7-12, are all consistent with spin-polarization tunneling in the former. To achieve more quantitative understanding of the spin-polarized tunneling in the manganite, we consider in the following subsection numerical simulations of the tunneling spectra under different conditions. 

\subsection{Simulations of the tunneling spectra}

To begin, it is instructive to consider the theoretically calculated DOS of both LCMO and Cr in the tunneling conductance. Specifically, the tunneling current ($I$) from Cr to LCMO as a function of the bias voltage ($V$) may be expressed by the following:
\begin{eqnarray}
I (V) &= G \sum _{\sigma , \sigma ^{\prime} = \alpha , \beta} \int d\omega {\cal N}_{\rm Cr} ^{\sigma} (\omega-eV) {\cal N}_{\rm LCMO} ^{\sigma ^{\prime}} (\omega) \nonumber\\
 &\times {\cal T} (\omega, H, \theta) \lbrack f_{\sigma}(\omega-eV) - f_{\sigma ^{\prime}}(\omega) \rbrack ,
\label{eq:IVall}
\end{eqnarray}
where $\omega$ denotes the quasiparticle energy relative to the Fermi level, $G$ is the LCMO sample conductance, $\sigma$ and $\sigma ^{\prime}$ refer to the spin-dependent energy bands (majority band: $\alpha$, minority band: $\beta$) of Cr and LCMO, respectively, $f_{\sigma} (\omega) = 1/ \lbrack \exp (\omega/k_B T) + 1 \rbrack$ is the Fermi-Dirac distribution function for $\sigma$-band, ${\cal N}_{\rm Cr} ^{\sigma} (\omega)$ and ${\cal N}_{\rm LCMO} ^{\sigma ^{\prime}}(\omega)$ are the spin-dependent DOS of Cr and LCMO, respectively, and ${\cal T} (\omega, H, \theta)$ represents the tunneling matrix of the STM junction that depends on the relative magnetization angle ($\theta$) of Cr and LCMO. Thus, the normalized tunneling conductance $(dI/dV)(V/I)$ can be determined from Eq.~(1) for given ${\cal N}_{\rm Cr} ^{\sigma} (\omega)$ and ${\cal N}_{\rm LCMO} ^{\sigma ^{\prime}} (\omega)$, provided that the relative spin configurations of Cr and LCMO are known and that the tunneling matrix is nearly independent of energy. 

\begin{figure*}
  \centering
  \includegraphics[width=5.2in]{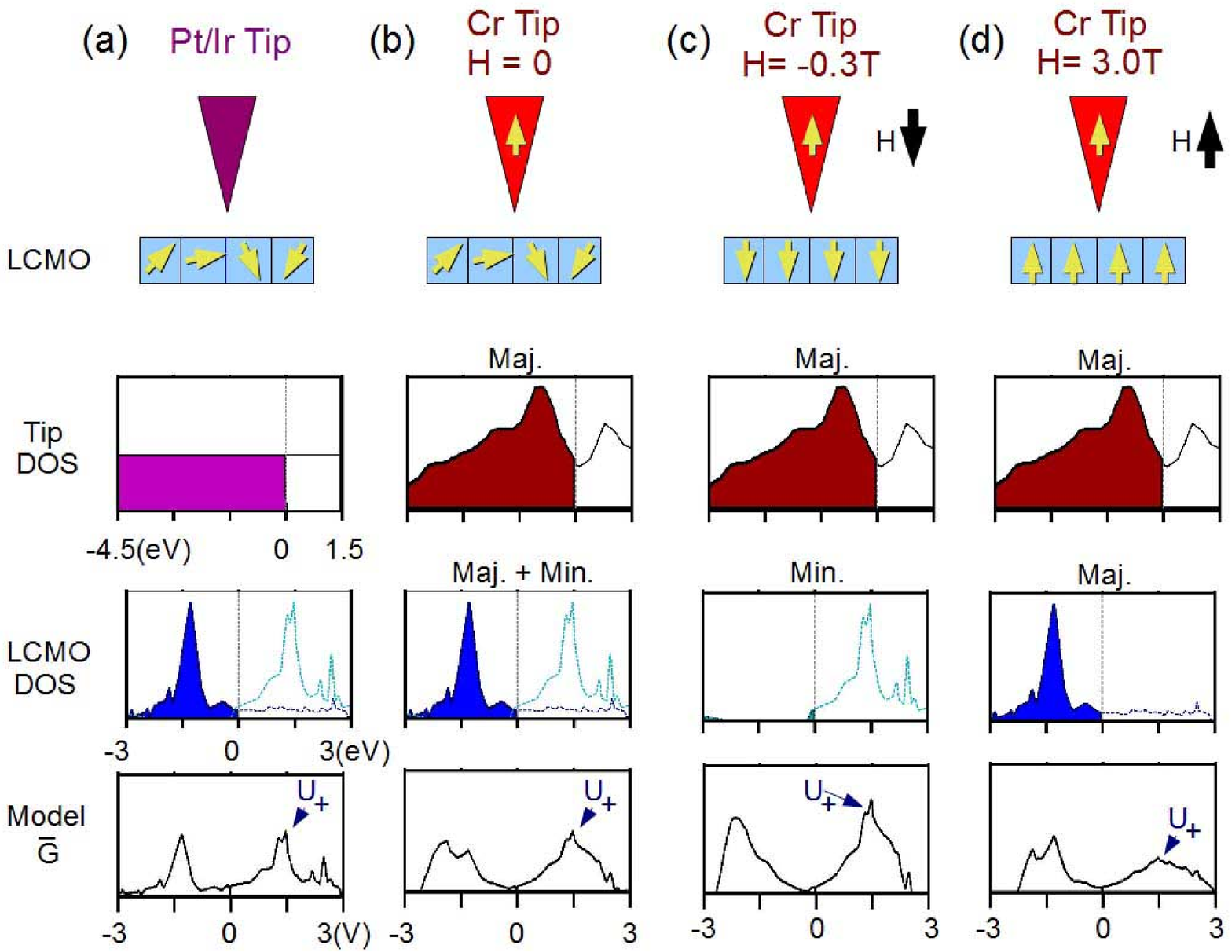}
\caption{(Color online) Effect of magnetic field and spin-polarized tunneling on the resulting tunneling spectra at $T \ll T_C ^{\rm LCMO}$: Schematic illustrations of STS measurements with (a) a Pt/Ir tip at $H = 0$, showing spin-degenerate tunneling currents into ferromagnetic LCMO with randomly oriented magnetic domains; (b) a Cr-coated tip at $H = 0$, showing spin-polarized tunneling currents into ferromagnetic LCMO with randomly oriented magnetic domains; (c) a Cr-coated tip at $H = -0.3$ T, showing spin-polarized tunneling currents into ferromagnetic LCMO with magnetic domains antiparallel to the spin polarization of the tunneling currents; (d) a Cr-coated tip at $H = 3.0$ T, showing spin-polarized tunneling currents into ferromagnetic LCMO with magnetic domains parallel to the spin polarization of the tunneling currents. The energy dependent DOS of the Cr-coated tip and that of the LCMO sample relevant to each spin configuration are shown in the second and third rows, respectively. We have also assumed in (a) that the DOS of Pt/Ir is a constant over the energy range of our consideration. Finally, the calculated tunneling conductance for each configuration is shown in the bottom row, which clearly illustrates maximum tunneling conductance $\bar{G}$ at $\omega = U_+$ for the anti-parallel spin configuration at $H = -0.3$ T.}
\label{fig14}
\end{figure*}

It is reasonable to assume that the magnetization of the Cr-coated tip follows the tip geometry for all temperatures of the experiments and is therefore fixed and approximately perpendicular to the plane of the LCMO sample. Therefore, at $H = 0$ the angle $\theta$ follows the spatial variation of the magnetization from one magnetic domain to another in the LCMO sample. On the other hand, for magnetic field opposite to the spin polarization of the Cr-coated tip and with a magnitude satisfying the condition $H_C ^{\rm LCMO} < |H| < H_C ^{\rm Cr}$, the net magnetization for the spin-polarized tunneling currents and the magnetization of each magnetic domain of LCMO are antiparallel, so that the selection rules for spins yield ${\cal T} (\omega, H, \theta = 0) = 0$, and for positive bias ($\omega = eV > 0$) the spin-polarized tunneling involves primarily electron tunneling (because hole tunneling is suppressed) from occupied Cr-majority band to empty LCMO-minority band and from occupied Cr-minority band to empty LCMO-majority band. Hence, for anti-parallel Cr and LCMO magnetizations, the normalized tunneling conductance $\bar{G}_{\uparrow \downarrow} (V)$ may be approximated by the following joint density of states:
\begin{equation}
\bar{G}_{\uparrow \downarrow} (V) \sim {\cal N}_{\rm Cr} ^{\alpha} (0) {\cal N}_{\rm LCMO} ^{\beta} (\omega) + {\cal N}_{\rm Cr} ^{\beta} (0) {\cal N}_{\rm LCMO} ^{\alpha} (\omega).
\label{eq:IVanti}
\end{equation}

Similarly, the normalized tunneling conductance for parallel magnetizations involves primarily electron tunneling from occupied Cr-majority band to empty LCMO-majority band and from occupied Cr-minority band to empty LCMO-minority band. Therefore, in the case of positive bias ($\omega = eV > 0$) we have ${\cal T} (\omega, H, \theta = \pm \pi) = 0$, and the normalized tunneling conductance $\bar{G}_{\uparrow \uparrow} (V)$ becomes:
\begin{equation}
\bar{G}_{\uparrow \uparrow} (V) \sim {\cal N}_{\rm Cr} ^{\alpha} (0) {\cal N}_{\rm LCMO} ^{\alpha} (\omega) + {\cal N}_{\rm Cr} ^{\beta} (0) {\cal N}_{\rm LCMO} ^{\beta} (\omega).
\label{eq:IVpara}
\end{equation}

Given Eqs.~(2) and (3) and the bandstructure calculations for ${\cal N}_{\rm Cr} ^{\alpha , \beta} (\omega)$ and ${\cal N}_{\rm LCMO} ^{\alpha , \beta} (\omega)$ illustrated in Fig.~13, the energy-dependent spin-polarized tunneling conductance may be derived. For various configurations under both regular and spin-polarized tunneling, as schematically illustrated in Figs.~14(a)-(d), the resulting tunneling spectra are calculated and shown in the bottom row of Fig.~14. Apparently the tunneling conductance at $\omega = U_+$ is significantly higher for the anti-parallel spin configuration between the Cr-coated tip and LCMO than that for the parallel configuration, consistent with the empirical histograms of the tunneling conductance shown in Fig.~8(d). Hence, the field-dependent tunneling spectra of LCMO obtained with a Cr-coated tip in the high-bias limit may be fully accounted for by spin-polarized tunneling in a spin-valve configuration.

\begin{figure}
  \centering
  \includegraphics[width=3.5in]{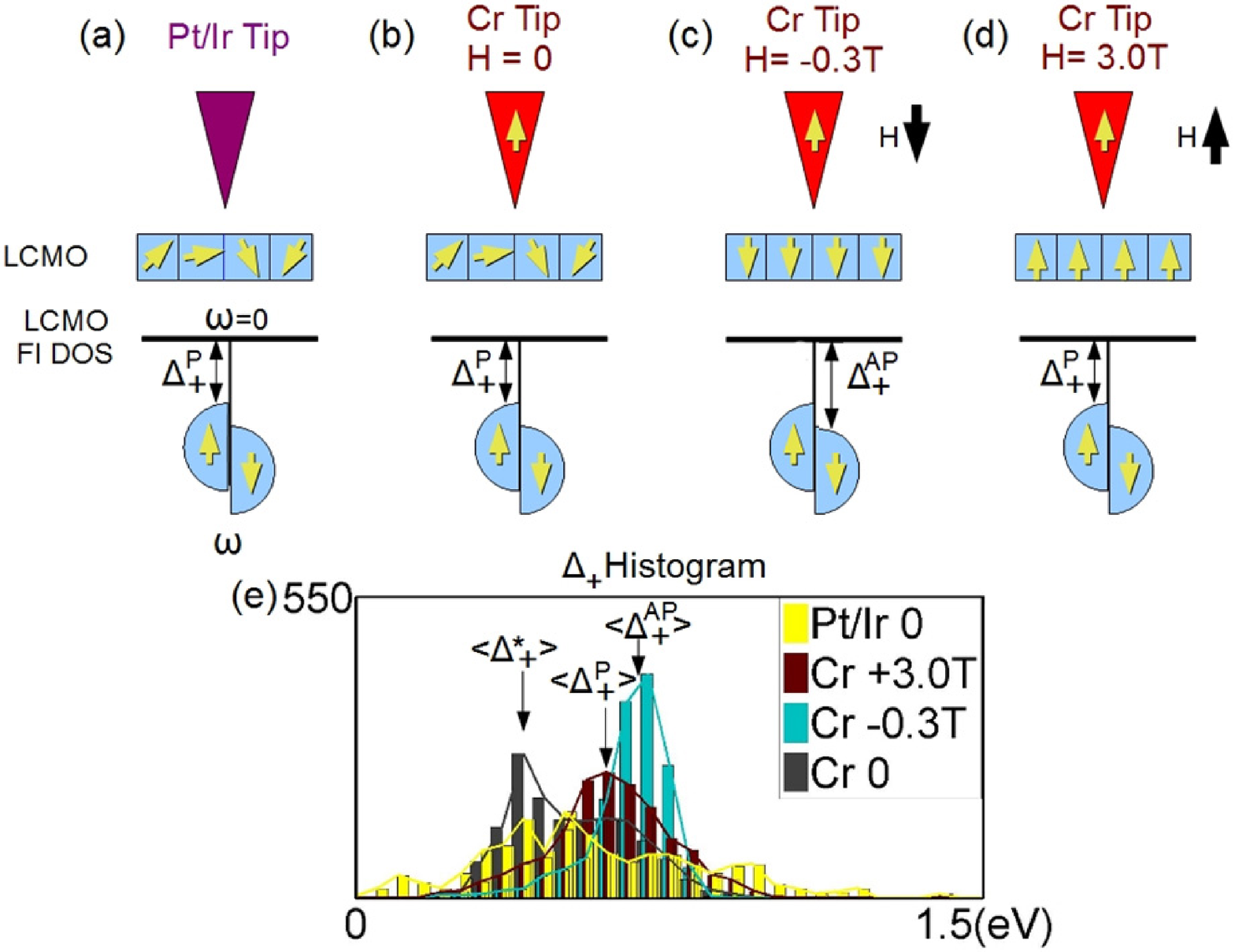}
\caption{(Color online) Effect of magnetic field and spin-polarized tunneling on the surface ferromagnetic insulating phase at low temperatures: Schematic energy ($\omega$) vs. density of states (DOS) plots of the surface ferromagnetic insulating phase in LCMO thin films, showing an energy gap $\Delta _+ ^{\rm P}$ relative to the Fermi level for the majority band and a larger energy gap $\Delta _+ ^{\rm AP}$ for the minority band. (a) For Pt/Ir tip at $H = 0$, the tunneling gap is determined by the minimum gap $\Delta _+ ^{\rm P} $ in addition to the presence of a pseudogap at $\Delta _+ ^{\ast}$ due to the electronic heterogeneity in the manganites; (b) For Cr-coated tip at $H = 0$, the tunneling gap $\Delta$ satisfies $\Delta _+ ^{\rm P} \le \Delta _+ \le \Delta _+ ^{\rm AP}$ in addition to a pseudogap at $\Delta _+ ^{\ast}$; (c) For Cr-coated tip at $H = -0.3$ T, the tunneling gap $\Delta _+$ is largely determined by the maximum gap $\Delta _+ ^{\rm AP}$, with the pseudogap mostly suppressed due to vanishing electronic heterogeneity in the presence of finite fields; (d) For Cr-coated tip at $H = 3.0$ T, the tunneling gap $\Delta _+$ is largely determined by the minimum gap $\Delta _+ ^{\rm P}$, and the pseudogap is also largely suppressed as in (c). The scenario described from (a) to (d) is consistent with the corresponding spin polarization- and field-dependent histograms of the gap shown in (e).}
\label{fig15}
\end{figure}

Next, we estimate the degree of spin polarization associated with the tunneling currents from the Cr-coated STM tip by the following consideration.~\cite{Bode03} Generally in a spin-valve configuration, the difference in the device tunneling conductance between the parallel and antiparallel magnetic configurations for the two ferromagnetic layers separated by an insulating barrier can be expressed by the following tunneling conductance ratio:~\cite{Bode03}
\begin{equation}
R_G (\omega) = \frac{\bar{G} _{\uparrow \uparrow} (\omega) - \bar{G} _{\uparrow \downarrow} (\omega)}{\bar{G} _{\uparrow \downarrow} (\omega)},
\label{eq:RG}
\end{equation}
where $\bar{G}_{\uparrow \uparrow}$ and $\bar{G}_{\uparrow \downarrow}$ represent the respective conductance in the parallel and antiparallel configurations. The energy-dependent tunneling conductance ratio in Eq.~(\ref{eq:RG}) may be further related to the bias-dependent spin-polarizations $P_1$ and $P_2$ of the two ferromagnetic layers as follows:~\cite{Bode03}
\begin{equation}
R_G (\omega) = \frac{2 P_1 (\omega) P_2 (\omega)}{1- P_1 (\omega) P_2 (\omega)}.
\label{eq:SP12}
\end{equation}
The aforementioned planar spin-valve configuration may be generalized to the case of spatially resolved SP-STM conductance maps at $\omega = \langle U_+ \rangle$ by integrating the quantities in the numerator and denominator in Eq.~(\ref{eq:SP12}) over the entire scanned area. As illustrated in Fig.~14, we may assign the normalized conductance values taken at $H = -0.3$ T to the antiparallel magnetic configuration and those taken at $H = 3.0$ T to the parallel magnetic configuration. Inserting the spatially resolved conductance $\bar{G}(\omega = \langle U_+ \rangle, x, y)$ into Eq.~(\ref{eq:SP12}), we obtain 
\begin{eqnarray}
\frac{\int \int dx dy \left[ \bar{G} _{H = 3 \rm T} (\langle U_+ \rangle, x, y) - \bar{G} _{H = -0.3 \rm T} (\langle U_+ \rangle, x, y) \right]}{\int \int dx dy \quad \bar{G} _{H = -0.3 \rm T} (\langle U_+ \rangle, x, y)} \nonumber \\
= \frac{2 P_{\rm Cr} (\omega) P_{\rm LCMO} (\omega)}{1- P_{\rm Cr} (\omega) P_{\rm LCMO} (\omega)}, \qquad \qquad \qquad \qquad \qquad \qquad \qquad 
\label{eq:SP-Cr}
\end{eqnarray}
where $P_{\rm Cr}$ is the spin polarization of the Cr tip, and $P_{\rm LCMO}$ is the polarization of the LCMO sample. From bandstructure calculations for LCMO with $x = 0.25$, $P_{\rm LCMO}$ is estimated at $\sim 61$\% at $\omega = \langle U_+ \rangle$.~\cite{Pickett96} Therefore, the resulting polarization of the Cr tip using Eq.~(\ref{eq:SP-Cr}) yields $P_{\rm Cr} \sim 15$\%.

\section{Discussion}

The simulations described in Section III are based on the assumption of regular and spin-polarized tunneling in bulk LCMO samples, and the simulated results are found to be consistent with the high-bias spectral behavior, indicating that tunneling currents from the Cr-coated tip may be considered as spin-polarized along a fixed direction for all temperatures and magnetic fields of our consideration. Next, we consider the low-bias spectral behavior under both regular and spin-polarized tunneling at 6 K, which provides useful information for the surface phase of LCMO. 

In the case of regular tunneling using Pt/Ir-tips, the histograms of the field-dependent insulating gap $\Delta _+$ shown in Fig.~10(e) reveal increasingly narrower distributions of gap values with increasing $|H|$, and the most probable value of $\Delta _+$ appears to independent of the magnetic field direction. In contrast, for spin-polarized tunneling using the Cr-coated tips, the histograms of the field-dependent insulating gap $\Delta _+$ shown in Fig.~7(e) reveal very different distributions from those shown in Fig.~10(e). Specifically, the zero-field $\Delta _+$ values obtained with a Cr-coated tip appear to be bi-modal, showing a sharper distribution peaking at the PG $\Delta _+ ^{\ast}$ and a boarder distribution centering around a gap value defined as $\Delta _+ ^{\rm P}$. On the other hand, for $H = -0.3$ T, the distribution of $\Delta _+$ becomes narrowly centering around a higher energy gap defined as $\Delta _+ ^{\rm AP}$ where $\Delta _+ ^{\rm AP} > \Delta _+ ^{\rm P}$, and the pseudogap features $\Delta _+ ^{\ast}$ appear to be strongly suppressed by magnetic field. In contrast, for $H = 3.0$ T, the distribution of $\Delta _+$ is largely concentrating around $\Delta _+ ^{\rm P}$ with a linewidth broader than that for $H = -0.3$ T, although the PG features are also strongly suppressed. 

The seemingly puzzling field-dependent $\Delta _+$ distributions shown in Figs.~7(e) and 10(e) may be understood by attributing the surface phase of LCMO to a ferromagnetic insulator (FI) as the result of one of the following two scenarios. The first scenario is to consider the LCMO sample as stoichiometric both in the bulk and near the surface, with the surface atomic layer terminated by insulating MnO$_2$.~\cite{Broussard97} In this context, the ferromagnetic nature of the sample is maintained throughout the sample, including the surface, so that the surface state behaves like a FI. Indeed, our previous optical studies of LCMO epitaxial films and single crystals~\cite{Boris97,Bazhenov98,Boris99} have revealed clear Drude-like optical conductivity below $T_C$, implying bulk metallic behavior in the ferromagnetic state. Hence, the bulk characteristics of our LCMO samples determined from measurements of the magnetization, transport and optical properties are all consistent with a metallic ferromagnetic phase,~\cite{Yeh97,Yeh97a,Yeh97b,Boris97,Bazhenov98,Boris99} whereas manifestation of a gapped state is only found in surface sensitive measurements such as the STM and XPS studies,~\cite{Wei97,Wei98,Seiro07,Moshnyaga06,Vasquez94,Vasquez96,Yeh97b} suggesting that only the surface of LCMO is insulating. Similar results from optical measurements have also been reported for cleaved $\rm La_{0.7}Sr_{0.3}MnO_3$ single crystals,~\cite{TakenakaK99} suggesting common metallic behavior in the bulk ferromagnetic phase of the manganites. 

On the other hand, a second senario that assumes a Ca-deficient surface layer may also account for the aforementioned experimental findings. Specifically, according to the phase diagram of $\rm La_{1-x}Ca_xMnO_3$,~\cite{Dagotto01,Algarabel03} the ground state of the manganites is a ferromagnetic insulator (FI) if the Ca doping level $x$ is less than 0.2. Although no experimental techniques known to date can determine the nano-scale local chemical compositions of a sample surface, it is conceivable that a Ca-deficient surface in the LCMO may result in a FI surface phase. However, the substantial Ca-deficiency required to yield a FI phase seemed incompatible with the averaged slight Ca-deficiency (Ca = $0.26 \pm 0.02$) from the XPS studies of LCMO.~\cite{Vasquez96} Overall, the gapped nature of the LCMO surface phase and the dependence of $\Delta _+$ on magnetic fields and on spin polarization are all consistent with a insulating surface phase differing from the metallic bulk. We further note that for LSMO/STO manganite superlattices (LSMO: $\rm La_{1-x}Sr_xMnO_3$, STO: $\rm SrTiO_3$), stoichiometric variations between the bulk LSMO and the interface of LSMO/STO have also been reported.~\cite{MaJX09,ZeniaH07} 

To examine the validity of a surface FI phase, we consider the field-dependent spectra under spin-polarized tunneling to the sample surface. If $H = 0$, the magnetization in the surface FI domains should be randomly oriented, so that the effective insulating gap values determined by spin-polarized tunneling currents from the Cr-coated tip are expected to range from the minimum gap $\Delta _+ ^{\rm P}$ to the maximum gap $\Delta _+ ^{\rm AP}$, as schematically illustrated in Figs.~15(a)-(b). In addition, a PG at a lower energy $\Delta _+ ^{\ast}$ should be present due to the electronic heterogeneity of the manganites at $H = 0$.~\cite{Moreo99,Moreo00} Similar observation is also expected for regular tunneling at $H = 0$ although the gap distribution is expected to be broader due to lack of spin selectivity. Indeed both regular and spin-polarized tunneling spectra at $H = 0$ are consistent with the empirical gap histograms shown in Fig.~15(e). On the other hand, for $H = -0.3$ T we expect the FI domains to be aligned so that a tunneling current with spin-polarization opposite to the LCMO surface magnetization results in a maximum insulating gap $\Delta _+ ^{\rm AP}$, as indicated in Fig.~15(c) and experimentally confirmed by the gap histogram in Fig.~15(e). Finally, for $H = 3.0$ T the spin-polarized tunneling currents are parallel to the magnetization of LCMO as illustrated in Fig.~15(d). Therefore, the insulating gap becomes $\Delta _+ ^{\rm P}$, which is the same as that found for regular tunneling currents and is smaller than the gap $\Delta _+ ^{\rm AP}$. The phenomenon of a larger energy gap for anti-parallel spin-polarized tunneling than that for parallel spin-polarized tunneling is known as the spin filtering effect, which is consistent with the empirical gap histogram shown in Fig.~15(e).  

On the other hand, the prominent presence of the PG phenomenon in the case of spin-polarized tunneling at $H = 0$ appears to diminish rapidly with increasing field, as shown in Fig.~15(e). While the PG energy $\Delta _+ ^{\ast}$ found at $H = 0$ exhibits little temperature dependence as shown in Figs.~7(f) and 10(f), the strong suppression of PG under finite magnetic fields and the occurrence of PG phenomena ($\gamma$-type spectra) at $H = 0$ primarily along the boundaries between phase separated regions (Fig.~3(d)) all implies that the PG phenomena in the manganites originate from the inherent electronic heterogeneity.~\cite{Moreo99,Moreo00} Consequently, increasing carrier mobility and magnetic domain alignments upon the application of magnetic fields can reduce the spatially inhomogeneous electronic properties in the manganites, thereby suppressing the PG phenomena.

Finally, it is interesting to compare the PG phenomena in the manganites with those in the cuprate superconductors.~\cite{Yeh09,Yeh10} While the PG persists in both perovskite oxides well above their ordering ($i.e.$, superconducting for the cuprates and ferromagnetic for the manganites) transition temperatures in the absence of magnetic fields, the application of an external magnetic field has opposite effects on these two systems. That is, PG becomes strongly {\it enhanced} by magnetic fields in the cuprates,~\cite{Beyer09,Teague09} which is in stark contrast to the significant {\it suppression} of PG by magnetic fields in the manganites. Hence, the physical origin for the occurrence of PG phenomena appears to be fundamentally different for these two strongly correlated perovskite oxides, which also suggests that the appearance of PG alone in the cuprates is unlikely a sufficient condition for the occurrence of high-temperature superconductivity.~\cite{Yeh09,Yeh10}  

\section{Conclusion}

In conclusion, spatially resolved regular and spin-polarized tunneling spectra on the CMR manganite $\rm La_{0.7}Ca_{0.3}MnO_3$ (LCMO) has been studied systematically by means of STM as a function of the temperature and applied magnetic fields. The spatial evolution reveals apparent phase separations on a length scale of $\sim 10^2$ nm, and the spectral characteristics obtained under the spin-valve configurations can be quantitatively accounted for by the spin-dependent joint density of states of the LCMO and the Cr-coated STM tip. The physical origin for a low-energy insulating gap detected at low temperatures has been investigated and attributed to a ferromagnetic insulating surface phase, which is consistent with the XPS findings of an insulating MnO$_2$ surface plane. Finally, spatially varying pseudogap phenomena with a nearly temperature independent pseudogap value have been observed in the zero-field tunneling spectra. The pseudogap phenomena appear primarily along the boundaries of phase separated regions and become strongly suppressed by applied magnetic fields when the tunneling spectra of LCMO become highly homogeneous. These results are consistent with the notion that the occurrence of pseudogap is associated with the electronic heterogeneity of the manganites. 

\begin{acknowledgments}
This research was supported jointly by the National Science Foundation through the Center of Science and Engineering of Materials (CSEM) at Caltech and the Kavli Foundation through the Kavli Nanoscience Institute (KNI) at Caltech. The SQUID data were taken at the Beckman Institute at Caltech. We thank Dr. Richard P. Vasquez for information about his XPS measurements on our LCMO samples, and Marcus L. Teague, Renee T.-P. Wu and Nils Asplund for their technical assistance. 
\end{acknowledgments}


\begin{thebibliography}{35}

\bibitem{Wolf01}
S. A. Wolf, D. D. Awschalom, R. A. Buhrman, J. M. Daughton, S. von Molnar, M. L. Roukes, A. Y. Chtchelkanova, D. M. Treger, Science {\bf 294}, 1488 (2001).

\bibitem{GuptaJA01}
J. A. Gupta, R. Knobel, N. Samarth and D. D. Awschalom, Science {\bf 292}, 2458 (2001).

\bibitem{Ramirez97}
A. P. Ramirez, J. Phys.: Condens. Matter {\bf 9}, 8171 (1997).

\bibitem{Coey99}
J. M. D. Coey, M. Viret and S. von Molnar, Adv. Phys. {\bf 48}, 167 (1999).

\bibitem{Kusters89}
R. M. Kusters {\it et al}, Physica B {\bf 155}, 362 (1989).

\bibitem{vonHelmolt93}
R. von Helmolt {\it et al}, Phys. Rev. Lett. {\it 71}, 2331 (1993).

\bibitem{Jin94}
S. Jin, T. H. Tiefel, M. McCormack, R. A. Fastnacht, R. Ramesh and L. H. Chen, Science {\bf 264}, 413 (1994).

\bibitem{Xiong95}
G. C. Xiong, Q. Li, H. L. Ju, S. N. Mao, L. Senapati, X. X. Xi, R. L. Greene and T. Venkatesan, Appl. Phys. Lett. {\bf 66}, 1427 (1995). 

\bibitem{Dagotto01}
E. Dagotto, T. Hotta and A. Moreo, Phys. Rep. {\bf 344}, 1 (2001).

\bibitem{Pickett96}
W. E. Pickett and D. J. Singh, Phys. Rev. B {\bf 53}, 1146 (1996).

\bibitem{Satpathy96}
S. Satpathy, Z. S. Popovic and F. R. Vukajlovic, Phys. Rev. Lett. {\bf 76}, 960 (1996)

\bibitem{Fath99}
M. F$\ddot{a}$th, S. Freisem, A. A. Menovsky, Y. Tomioka, J. Aarts, J. A. Mydosh, Science {\bf 285}, 1540 (1999).

\bibitem{Renner00}
Ch. Renner, G. Aeppli, B.-G. Kim, Yeong-Ah Soh and S.-W. Cheong, Nature {\bf 416}, 518 (2000)

\bibitem{Becker02}
T. Becker, C. Streng, Y. Luo,V. Moshnyaga, B. Damaschke, N. Shannon and K. Samwer, {\it Phys. Rev. Lett.} {\bf 89}, 237203 (2002)

\bibitem{Algarabel03}
P. A. Algarabel, J. M. De Teresa, J. Blasco, M. R. Ibarra, Cz. Kapusta, M. Sikora, D. Zajac, P. C. Riedi and C. Ritter, Phys. Rev. B {\bf 67}, 134402 (2003).

\bibitem{Aruta09}
C. Aruta, G. Ghiringhelli, V. Bisogni, L. Braicovich, N. B. Brookes, A. Tebano and G. Balestrino, Phys. Rev. B {\bf 80}, 014431 (2009). 

\bibitem{Wei97}
J. Y. T. Wei, N.-C. Yeh and R. P. Vasquez, Phys. Rev. Lett. {\bf 79}, 5150 (1997). 

\bibitem{Wei98}
J. Y. T. Wei, N.-C. Yeh, R. P. Vasquez and A. Gupta, J. App. Phys. {\bf 83}, 7366 (1998)

\bibitem{Seiro07}
S. Seiro, Y. Fasano, I. Maggio-Aprile, O. Kuffer and �. Fischer, J. Mag. Mag. Mat. {\bf 310}, E243 (2007)

\bibitem{Moshnyaga06}
V. Moshnyaga, L. Sudheendra, O. I. Lebedev, S. A. Koster, K. Gehrke, O. Shapoval, A. Belenchuk, B. Damaschke, G. van Tendeloo and K. Samwer, Phys. Rev. Lett. {\bf 97}, 107205 (2006).

\bibitem{Yeh97}
N.-C. Yeh, R. P. Vasquez, D. A. Beam, C.-C. Fu, J. Huynh and G. Beach, J. Phys.: Condens. Matter {\bf 9}, 3713 (1997).

\bibitem{Yeh97a}
N.-C. Yeh, C. C. Fu, J. Y. T. Wei, R. P. Vasquez, J. Huynh, S. M. Maurer, D. A. Beam and G. Beach, J. Appl. Phys. {\bf 81}, 5499 (1997).

\bibitem{Broussard97}
P. R. Brousaard, S. B. Qadri, V. M. Browning and V. C. Cestone, Appl. Surf. Sci. {\bf 115}, 80 (1997).

\bibitem{Moreo99}
A. Moreo, S. Yunoki, and E. Dagotto, Phys. Rev. Lett. {\bf 83}, 2773 (1999).

\bibitem{Moreo00}
A. Moreo, M. Mayr, A. Feiguin, S. Yunoki and E. Dagotto, Phys. Rev. Lett. {\bf 84}, 5568 (2000).

\bibitem{Dessau98}
D. S. Dessau, T. Saitoh, C.-H. Park, Z.-X. Shen, P. Villella, N. Hamada, Y. Moritomo and Y. Tokura, Phys. Rev. Lett. {\bf 81}, 192 (1998).

\bibitem{Yeh09}
N.-C. Yeh and A. D. Beyer, Int. J. Mod. Phys. B {\bf 23}, 4543 (2009).

\bibitem{Yeh10}
N.-C. Yeh, A. D. Beyer, M. L. Teague, S.-P. Lee, S. Tajima and S. I. Lee, J. Supercond. Nov. Magn. {\bf 23}, 757 (2010).

\bibitem{Bode03}
M. Bode, Rep. Prog. Phys. {\bf 66}, 523 (2003).

\bibitem{Czerner09}
M. Czerner, G. Rodary, S. Wedekind, D. V. Fedorov, D. Sander, I. Mertig, and J. Kirschner, J. Mag. Mag. Mat. {\bf 322}, 1416 (2009).

\bibitem{Alvarado95}
S. F. Alvarado, Phys. Rev. Lett. {\bf 75}, 513 (1995).

\bibitem{Alvarado92}
S. F. Alvarado and P. Renaud, Phys. Rev. Lett. {\bf 68}, 1387 (1992).

\bibitem{Meier08}
F. Meier, L. Zhou, J. Wiebe and R. Wiesendanger, Science {\bf 320}, 82 (2008).

\bibitem{Mitra05}
J. Mitra, M. Paranjape, and A. K. Raychaudhuri, N. D. Mathur and M. G. Blamire, Phys. Rev. B {\bf 71}, 094426 (2005).

\bibitem{Yeh97b}
N.-C. Yeh, R. P. Vasquez, J. Y. T. Wei, C. C. Fu, G. Beach, J. Huynh, A. V. Samoilov, A. V. Boris, N. N. Kovaleva, and A. V. Bazhenov, in ``Epitaxial Oxide Thin Films -- III'', Mat. Res. Soc. Sym. Proc. {\bf 474}, 145 (1997).

\bibitem{ChoiJ99}
J. Choi, J. Zhang, S.-H. Liou, P. A. Dowben and E. W. Plummer, Phys. Rev. B {\bf 59}, 13453 (1999).

\bibitem{Vasquez94}
R. P. Vasquez, J. Electron Spectros. Rel. Phenomena {\bf 66}, 209 (1994); {ibid.} {\bf 66}, 241 (1994).

\bibitem{Vasquez96}
R. P. Vasquez, Phys. Rev. B {\bf 54}, 14938 (1996); also R. P. Vasquez, private communications.

\bibitem{correlation}
The degree of cross correlation between the $U_+$ and $\Delta _+$ maps may be quantified by considering the cross-correlation function $C(\textbf{R})$ defined as follows:
\begin{equation}
\label{correlation}
C(\textbf{R}) \equiv \frac{\int _S d^2 r \lbrack f(\textbf{r}) - \langle f \rangle \rbrack \lbrack g(\textbf{r}+\textbf{R}) - \langle g \rangle \rbrack}{\sqrt{A_{ff} (0) A_{gg} (0)}}.
\end{equation}
Here $f(\textbf{r})$ and $g(\textbf{r})$ represent the $U_+$ and $\Delta _+$ values at position $(\textbf{r})$, $\langle f \rangle$ and $\langle g \rangle$ are the mean values over the sample area $S$, and $A_{ff} (A_{gg})$ denotes the auto correlation function of $U_+$ ($\Delta _+$):
\begin{equation}
\label{eq:cross}
A_{ff} (\textbf{R}) \equiv \int d^2 r \lbrack f(\textbf{r}) - \langle f \rangle \rbrack \lbrack f(\textbf{r}+\textbf{R}) - \langle f \rangle \rbrack.
\end{equation}
The definition in Eq.~(\ref{eq:cross}) implies complete correlation for a value of $+1$, complete anti-correlation for $-1$, and no correlation for $0$. 

\bibitem{Boris97}
A. V. Boris, A. V. Bazhenov, N. N. Kovaleva, A. V. Samoilov, N.-C. Yeh, and R. P. Vasquez, J. Appl. Phys. {\bf 81}, 5756 (1997).

\bibitem{Bazhenov98}
A. V. Bazhenov, A. V. Boris, N. N. Kovaleva, A. V. Samoilov, N.-C. Yeh, and R. P. Vasquez, Inst. Phys. C {\bf 160}, 429 (1998).

\bibitem{Boris99}
A. V. Boris, N. N. Kovaleva, A. V. Bazhenov, P. J. M. van Bentum, Th. Rasing, S.-W. Cheong, A. V. Samoilov, and N.-C. Yeh, Phys. Rev. B {\bf 59}, R697 (1999).

\bibitem{TakenakaK99}
K. Takenaka, K. Iida, Y. Sawaki, S. Sugai, Y. Moritomo and A. Nakamura, J. Phys. Soc. Jpn. {\bf 68}, 1828 (1999).

\bibitem{MaJX09}
J. X. Ma, X. F. Liu, T. Lin, G. Y. Gao, J. P. Zhang, W. B. Wu, X. G. Li and J. Shi, Phys. Rev. B {\bf 79}, 174424 (2009).

\bibitem{ZeniaH07}
H. Zenia, G. A. Gehring and W. M. Temmerman, New J. Phys. {\bf 9}, 105 (2007).

\bibitem{Beyer09}
A. D. Beyer, M. S. Grinolds, M. L. Teague, S. Tajima and N.-C. Yeh, Europhys. Lett. {\bf 87}, 37005 (2009).

\bibitem{Teague09}
M. L. Teague, A. D. Beyer, M. S. Grinolds, S. I. Lee and N.-C. Yeh, Europhys. Lett. {\bf 85}, 17004 (2009).

\end{thebibliography}
\end{document}